\PassOptionsToPackage{unicode}{hyperref}
\PassOptionsToPackage{hyphens}{url}
\PassOptionsToPackage{dvipsnames,svgnames,x11names}{xcolor}

\documentclass[12pt]{article}

\RequirePackage{amsthm,amsmath,amsfonts,amssymb}
\RequirePackage{natbib}
\RequirePackage{graphicx}%% uncomment this for including figures
\usepackage{xcolor}
\usepackage{float}
\usepackage{outlines}
\usepackage[toc,page]{appendix}
\usepackage{verbatim}
\usepackage{algorithm2e} % for the algorithm box
\usepackage{algorithmic} % for the algorithm box
\usepackage{fix-cm}
\usepackage[T1]{fontenc}
\usepackage{newtxtext}
\usepackage{bm}
\usepackage[english]{babel}

\usepackage[T1]{fontenc}
\usepackage[utf8]{inputenc}
\usepackage{textcomp}

\RequirePackage[colorlinks,citecolor=blue,urlcolor=blue]{hyperref}%% uncomment this for coloring bibliography citations and linked URLs

\graphicspath{{../figures/}}

\newcommand{\by}{\boldsymbol{y}}
\newcommand{\btheta}{\boldsymbol{\theta}}
\newcommand{\bc}{\boldsymbol{c}}
\newcommand{\bmu}{\boldsymbol{\mu}}
\newcommand{\bSigma}{\boldsymbol{\Sigma}}

\newcommand{\bcm}{c_1,c_2,\ldots,c_{i-1}}
\newcommand{\bcmperm}[1]{c_{\sigma_1},\ldots,c_{\sigma_{#1}}}
\newcommand{\byperm}{\by_{\sigma_i}}

\newcommand{\fmly}{\boldsymbol{x}}

\newcommand{\crp}{_{CRP}}
\newcommand{\dfcrp}{_{DFCRP}}
\newcommand{\bsigma}{\boldsymbol{\sigma}}

\begin{document}

\def\spacingset#1{\renewcommand{\baselinestretch}
{#1}\small\normalsize}\spacingset{1}

\title{\bf Clustering Craters on the Moon with Dysfunctional Families}
\author{Nathan Weed \hspace{.2cm}\\
    Department of Statistics, Brigham Young University\\
    and \\
    Emily Castleton  \hspace{.2cm}\\
    Statistics, Los Alamos National Laboratory\\
    and \\
    Dave Osthus  \hspace{.2cm}\\
    Statistics, Los Alamos National Laboratory\\
    and \\
    Brian Weaver  \hspace{.2cm}\\
    Statistics, Los Alamos National Laboratory\\
    and \\
    Richard L. Warr  \hspace{.2cm}\\
    Department of Statistics, Brigham Young University\\
    }
  \maketitle

\newpage

\begin{abstract}
Summaries of craters on terrestrial bodies, such as the number and size distribution, are essential for understanding the history of the Solar System. Identifying craters, however, has not been automated and thus relies on expert crater-counters marking static images. \citet{r14} (hereafter R14) showed that, contrary to previously held assumptions, there exists large variability across expert crater-counters' identified crater lists. How best to combine identified crater lists across multiple experts for the purposes of learning about the Solar System is an open and consequential question. R14 combined identified crater lists via clustering through a modification of the popular DBSCAN clustering method. Their approach did not, however, make use of all the constraining information available nor did it provide an estimate of clustering uncertainty. To address the shortcomings of the DBSCAN method, we present a novel clustering approach that can combine multiple lists of identified objects of interest from the same image. The key innovation is incorporating a dysfunctional family constraint into the Bayesian non-parametric clustering approach, the Chinese restaurant process (CRP), which naturally takes into account information about the crater identifier. The dysfunctional family Chinese restaurant process (DFCRP) provides an estimate of clustering uncertainty. In this work, we provide guidance on hyperparameter specification, present a Gibbs sampler, and perform a simulation study to compare the performance of the DFCRP to the CRP. Finally, we apply the DFCRP to the crater identification problem of R14, comparing results, and also demonstrate the types of analyses that can be performed with posterior draws of cluster assignments.
\end{abstract}

\noindent%
{\it Keywords:} %3 to 6 keywords, that do not appear in the title
Bayesian nonparametric methods,
Chinese restaurant process,
cross-expert inconsistency,
constrained clustering,
uncertainty quantification
\vfill

\newpage
\spacingset{1.8} % DON'T change the spacing!

\section{Problem Background}
Counting craters on terrestrial bodies (i.e., ``Earth-like'' or rocky bodies) is essential for understanding the history of the Solar System.  The process of counting craters has not been automated by an algorithm: a human carefully examines an image (e.g., of the moon) and attempts to exhaustively identify the craters. This is typically done with the assistance of a software package where a circle is placed over the identified crater (\citet{r14}, hereafter R14). Various summaries of the identified craters are used to make scientific statements about the rocky body.  For example, the number of craters within a certain diameter range can estimate the age of the body based on lunar rock samples retrieved from the Apollo missions \citep{kcmceb13}.

\begin{figure}[!tb]
    \centering
    \includegraphics[width=284.2pt]{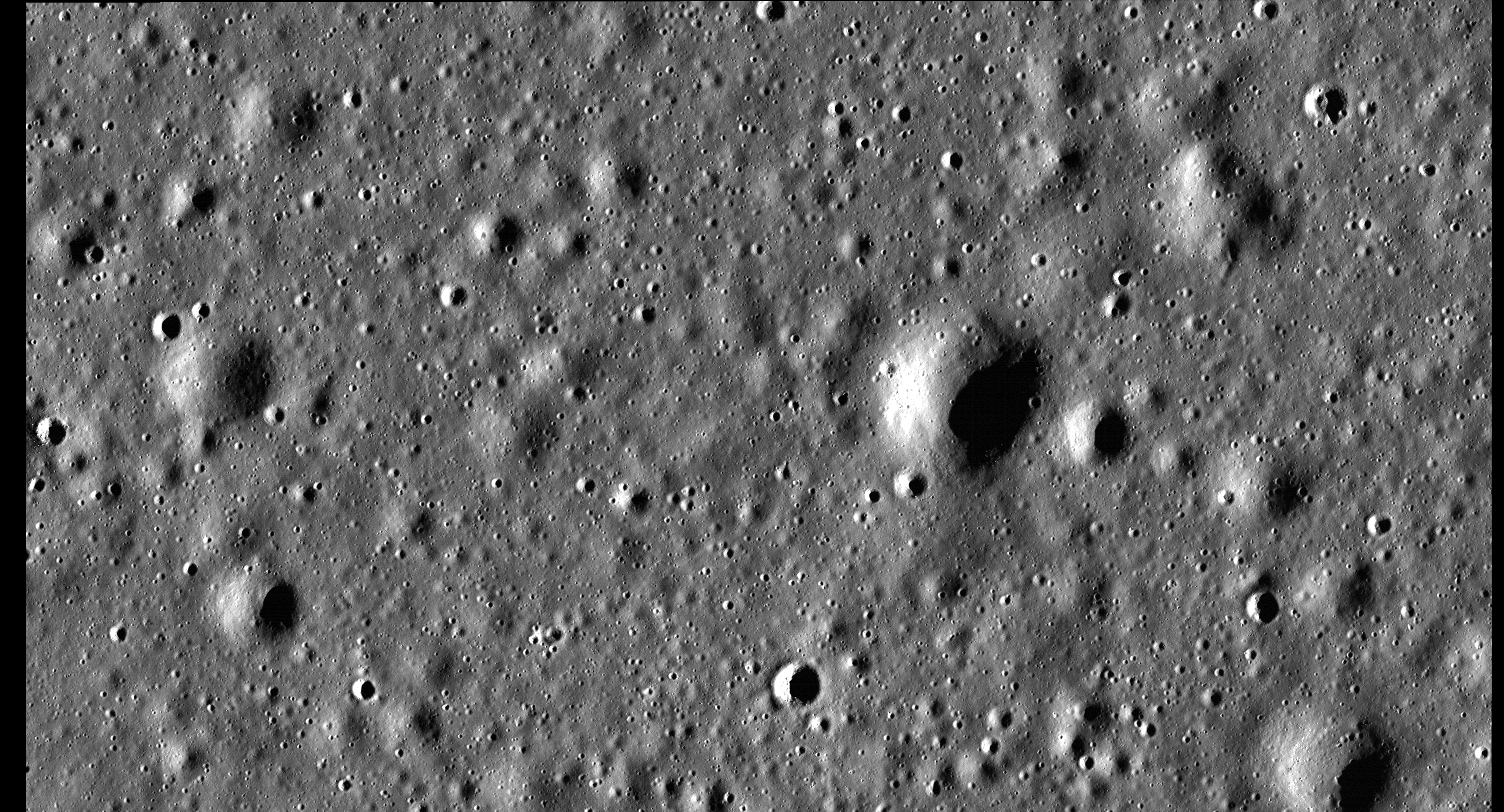}
    \caption{The 4107 $\times$ 2218 pixel image of the lunar highlands distributed to crater counters. The image is available in the supplementary data of \cite{r14}. For scale, the area of the moon shown in the image is roughly 2.43 miles (width) by 1.2 miles (height).}
    \label{fig:highlands}
\end{figure}

An assumption within the crater-counting community is that crater identification and crater size measurements are straightforward, and thus, crater counters are interchangeable (i.e., multiple crater counters examining the same image will  identify identical craters). R14 designed a study to test this assumption. An image of the lunar highlands, shown in Figure \ref{fig:highlands}, was distributed to eight crater counters. They were asked to identify all the craters in the image. For this analysis, we only considered craters with diameters greater than or equal to 18 pixels (px). Some of the eight crater counters used multiple software packages to count the craters, resulting in 11 unique lists. For simplicity, in this work, we will refer to each combination of human crater counter and software as an ``expert''. When combining crater counts across experts, R14 found substantial variability between assessments.

\begin{figure}[!tb]
    \centering
    \includegraphics[width=360pt,trim={0pt 0pt 0pt 0pt},clip]{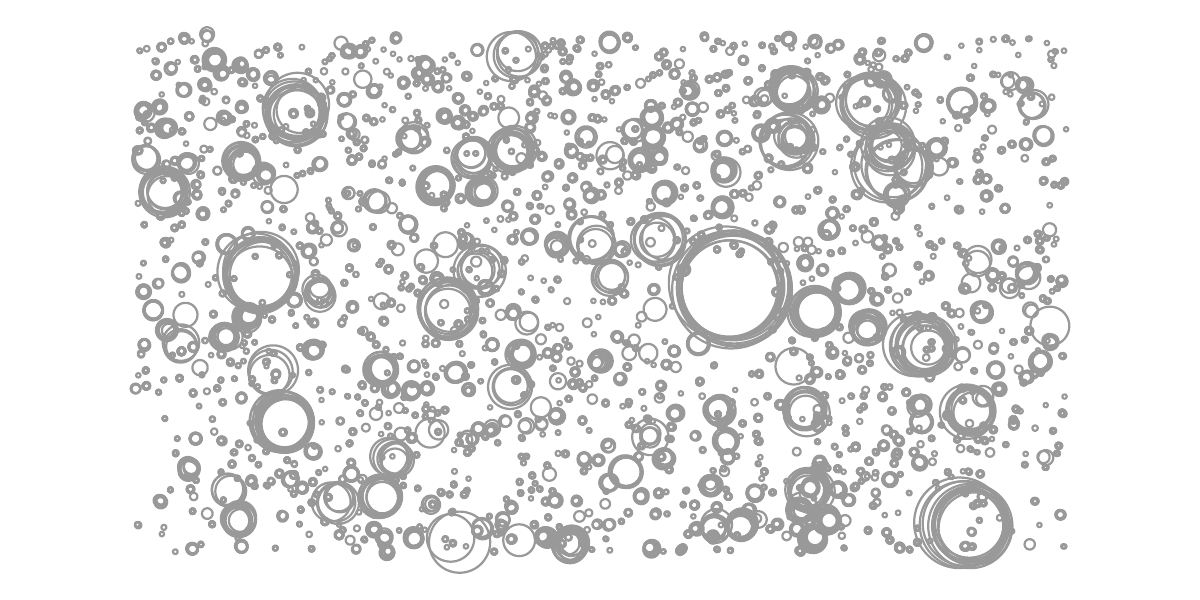}
    \caption{Observational craters from all 11 crater counting experts in the lunar highlands image (Figure \ref{fig:highlands}). The size of the plotting symbol is proportional to the crater's diameter.}
    \label{fig:data}
\end{figure}

All 9,517 identified craters, referred to here as \emph{observational craters}, from the 11 experts on the lunar highlands image are shown in Figure \ref{fig:data}, where the size of the plotting symbol is proportional to the crater's identified diameter. The number of observational craters by experts ranged from 636 to 1,197---a significant range. The average number of observational craters identified by an expert was 865. 

The large variability between experts' observational crater lists poses challenges for scientific learning. If the size and number of craters are in fact uncertain, then the results of the methods that rely on them are also uncertain. Furthermore, conclusions drawn from only one expert's observational crater list fails to account for the variability between experts and subsequent inference depends on which expert performed the crater counting. 

\begin{figure}[!tb]
    \centering
    \includegraphics[width=64.96pt,trim={1.8cm 1.8cm 1.8cm .8cm},clip]{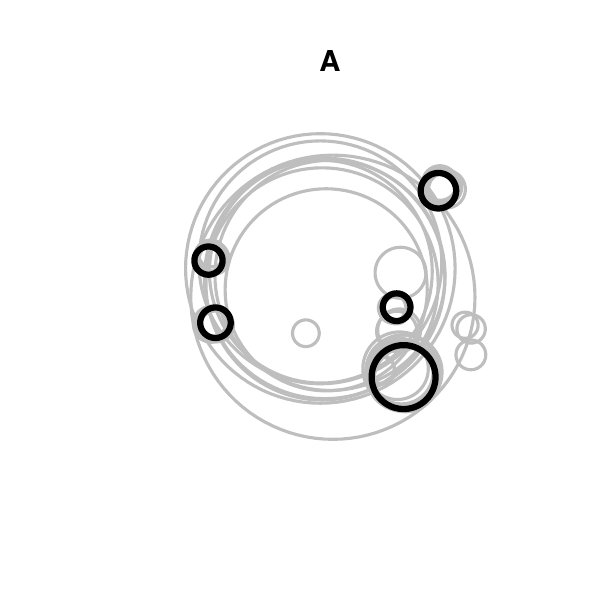}
    \includegraphics[width=64.96pt,trim={1.8cm 1.8cm 1.8cm .8cm},clip]{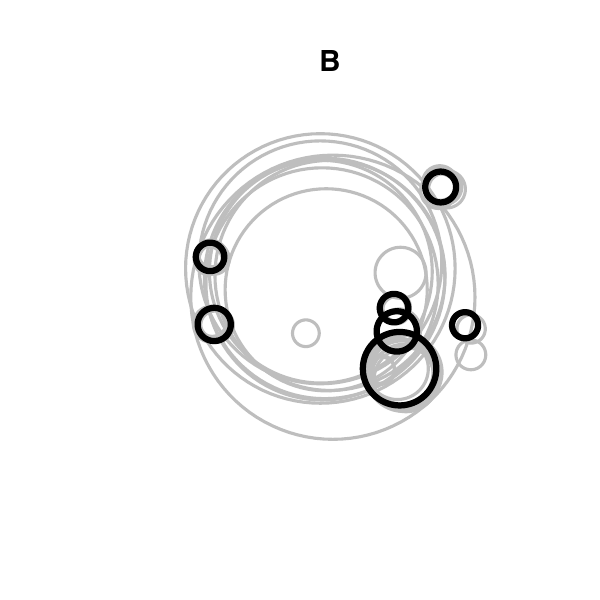}
    \includegraphics[width=64.96pt,trim={1.8cm 1.8cm 1.8cm .8cm},clip]{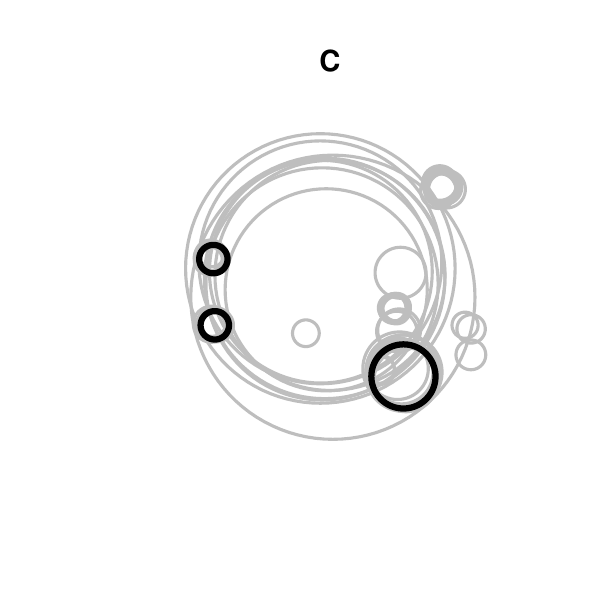}
    \includegraphics[width=64.96pt,trim={1.8cm 1.8cm 1.8cm .8cm},clip]{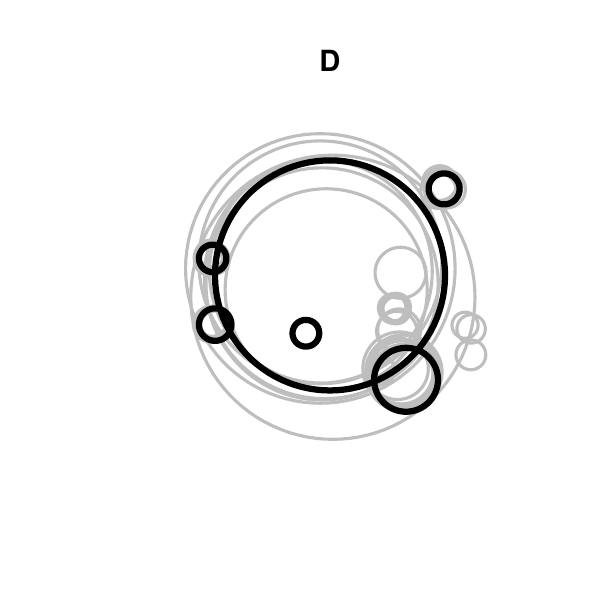}
    \includegraphics[width=64.96pt,trim={1.8cm 1.8cm 1.8cm .8cm},clip]{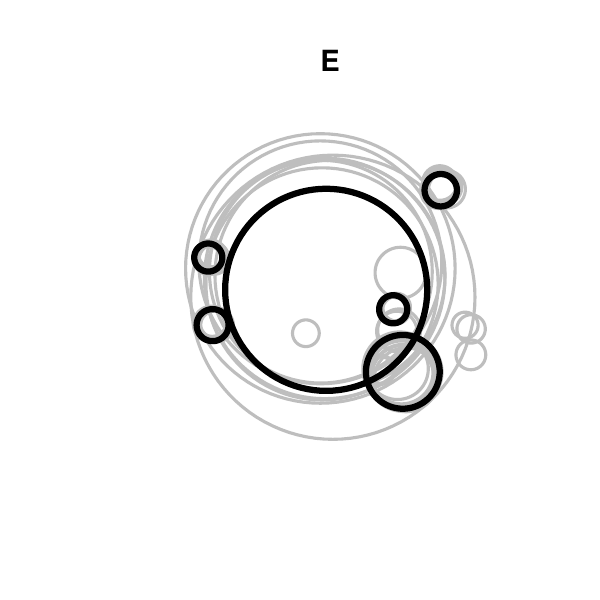}
    \includegraphics[width=64.96pt,trim={1.8cm 1.8cm 1.8cm .8cm},clip]{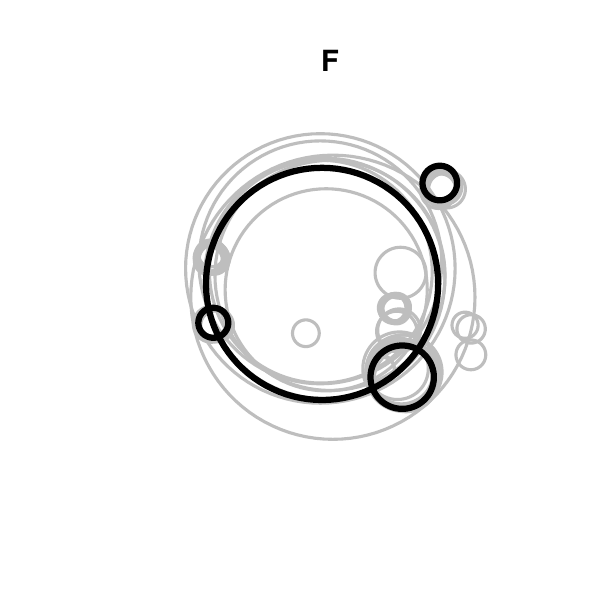}
    \includegraphics[width=64.96pt,trim={1.8cm 1.8cm 1.8cm .8cm},clip]{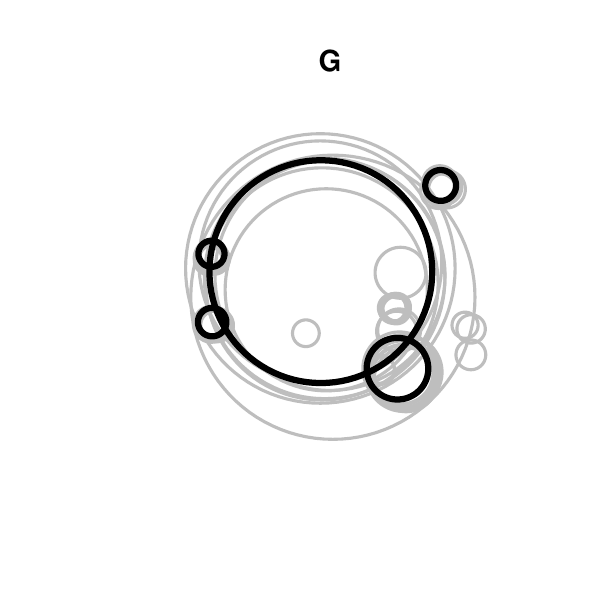}
    \includegraphics[width=64.96pt,trim={1.8cm 1.8cm 1.8cm .8cm},clip]{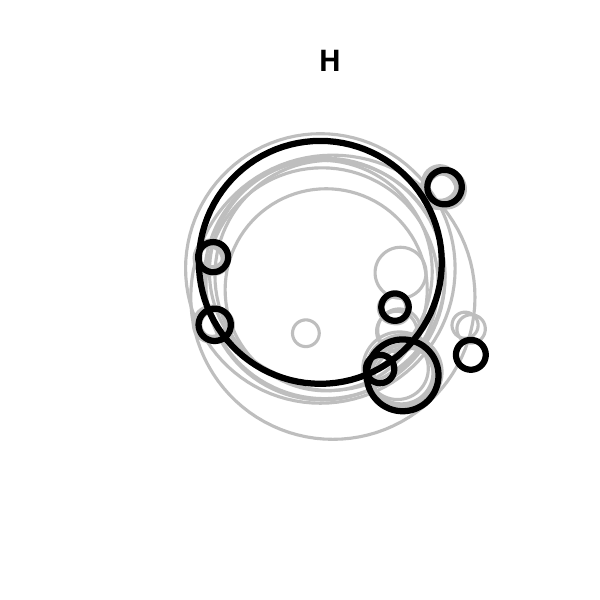}
    \includegraphics[width=64.96pt,trim={1.8cm 1.8cm 1.8cm .8cm},clip]{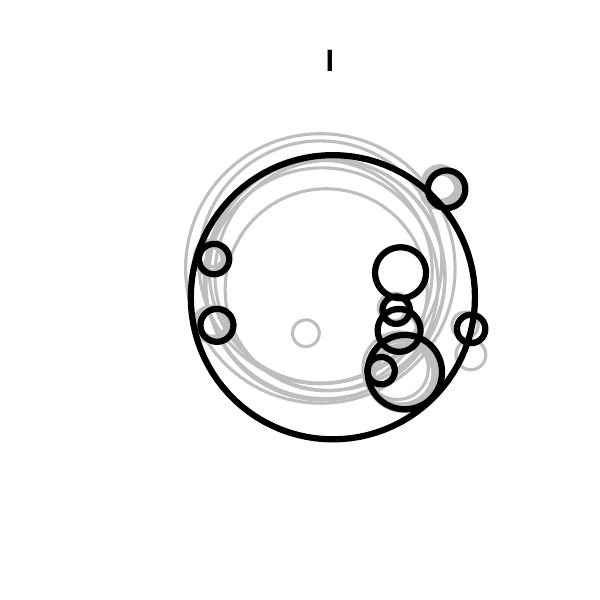}
    \includegraphics[width=64.96pt,trim={1.8cm 1.8cm 1.8cm .8cm},clip]{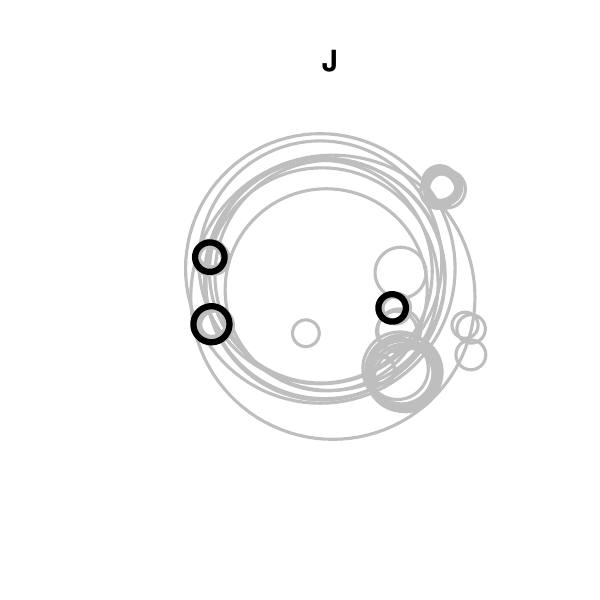}
    \includegraphics[width=64.96pt,trim={1.8cm 1.8cm 1.8cm .8cm},clip]{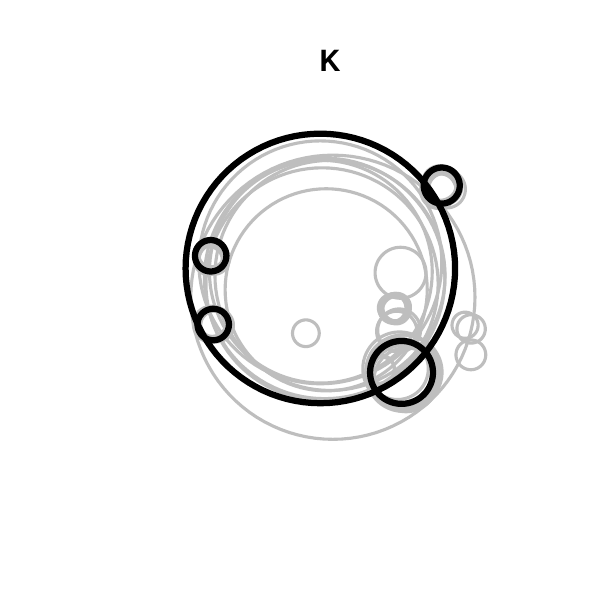}
    \caption{Observational craters identified by crater counters using specific software (experts). Black craters were identified by a specific expert, and gray craters were identified by the other experts. Observational craters show appreciable variability across experts. Experts C and J identified only three craters, while expert I identified 10.}
    \label{fig:problemarea}
\end{figure}

Acknowledging the large variability in crater identification across experts, the crater counting community is asking what is the best way estimate the number and size distribution of craters on the moon. R14 felt the best way to address these questions was via clustering. Clustering is an approach that can estimate the number and size distribution of craters on the moon by addressing two forms of variability in the data: crater identification and crater specification. Figure \ref{fig:problemarea} is a subimage of Figure \ref{fig:data} and illustrates both forms of variability. 

Crater identification variation is due to different experts identifying different craters. In Figure \ref{fig:problemarea}, this can be seen as the variation in the number of identified observational craters. For example, experts C and J only identified three craters in the subimage of Figure \ref{fig:problemarea} while expert I identified 10 craters. The second source of variability is crater specification. Crater counting software assumes that craters are perfect circles, so each observational crater is defined by three features:  latitude, longitude, and diameter. Crater specification variability is variability in the features. In Figure \ref{fig:problemarea}, experts D, E, F, G, H, I, and K identified a large crater, as indicated by a large circle. There is, however, variability in the three features between experts. For instance, expert E's large circle is smaller than expert I's large circle, raising the question, is expert E's large circle referring to the same crater as expert I's? Identifying which observational craters refer to the same underlying crater poses a challenge due to variability in both crater identification and specification. Clustering is an approach that can address these sources of variability.

R14 used a modification of a popular clustering algorithm, DBSCAN \citep{eksx96}, to group observational craters into clusters. The original DBSCAN algorithm requires the specification of two parameters, (1) a  minimum cluster size, and (2) a reachability parameter, which identifies points that can be clustered together with respect to distance between them. A modification was required to incorporate the crater diameter as a feature through the introduction of a second reachability parameter. Observational craters clustered together were interpreted as multiple observations of the same underlying crater. The minimum cluster size was set to five, where the size of the cluster was interpreted as the number of unique experts identifying the crater. All clusters of size five or more were referred to as \emph{consensus clusters} and assumed to be real craters. The number of consensus clusters was an estimate of the number of craters and the distribution of the consensus cluster diameter averages was an estimate of the size distribution of craters on the moon.

\iffalse
The interpretation that the cluster size equals the number of unique experts is predicated on an assumption--the assumption that multiple observational craters from the same expert cannot be assigned to the same cluster. As the DBSCAN clustering algorithm has no natural mechanism by which to enforce this constraint, this assumption is likely untrue. Assigning multiple observational craters to the same cluster, however, overrides the opinion of an expert that multiple, distinct craters exist. Furthermore, the DBSCAN algorithm does not provide an estimate of uncertainty on the cluster assignments, and thus no uncertainty of the total number of craters or the distribution of crater sizes.
\fi

We, like R14, believe combining information across multiple experts via clustering is a good approach. However, the clustering approach used by R14 failed to account for valuable clustering information provided by the experts themselves. In their DBSCAN clustering approach, R14 assumed cluster size equaled the number of unique experts that found the same underlying crater. However, there was no mechanism in place to ensure this interpretation. Placing multiple observational craters from the same expert in the same cluster and interpreting the cluster as a single real crater inappropriately overrides the opinion of the expert that multiple, distinct craters exist. Thus, R14 failed to account for the expert information source in their clustering. Rather than making this assumption, a clustering algorithm that incorporates information about experts as a constraint on the clustering assignments is more desirable. Furthermore, the DBSCAN algorithm does not provide an estimate of uncertainty on either the number of consensus clusters (i.e., craters) or the size distribution of cluster diameter averages (i.e., size distribution of craters). Given these estimates are based on highly variable counts from experts, an estimate on clustering uncertainty should be provided.

To address the shortcomings of R14, we develop a modification of the Chinese restaurant process (CRP) \citep{a85}, named the \emph{dysfunctional family Chinese restaurant process (DFCRP)}, and illustrate how it can be used to combine the crater identifications from multiple experts such that every cluster is composed of a unique set of experts. The DFCRP, as an extension of the CRP, also quantifies uncertainty on cluster assignment and functions thereof in a principled manner. 

The remainder of the this paper is organized as follows. Section \ref{sec:relatedwork} discusses related clustering work and motivates the development of the DFCRP. Section \ref{sec:crp} describes the traditional CRP.  The development and description of the DFCRP is presented in Section \ref{sec:dfcrp}. The joint specification of the DFCRP mixture model and hyperparameter selection are discussed in Section \ref{sec:dfcrpmm}. Posterior sampling is discussed in Section \ref{sec:posteriorsampling}. A simulation study is presented in Section \ref{sec:sim_study}. Section \ref{sec:full} details the application of the DFCRP to 11 expert-identified observational craters lists on the lunar highlands image. Finally, Section \ref{sec:conclusions} concludes the paper with a summary of findings and a discussion of future considerations.

\section{Related Work}
\label{sec:relatedwork}

There are multiple challenges to clustering observational craters for which a direct application of existing clustering methodologies is not straightforward. In this section, we evaluate $k$-means algorithms, DBSCAN (especially as used by R14), the constrained Dirichlet process model (C-DPM), and the CRP (along with its extensions). We compare these methods with our model, the DFCRP, using four criteria motivated by the nature of the data and the goals of the analysis.

Firstly, the number of clusters is not known \textit{a priori}, eliminating the possibility of using standard, simple clustering algorithms like $k$-nearest neighbors \citep{a92}, Lloyd's algorithm (i.e., $k$-means clustering) \citep{l82}, or spectral clustering \citep{njw02}, all of which require specifying $k$.

Next, we desire a clustering algorithm which can naturally incorporate  any number of features.  The modified DBSCAN method used by R14 was able to account for all three features (latitude, longitude, and crater diameter), but it required an extension that does not necessarily scale to more features. Further, this method has been known to suffer from the ``curse of dimensionality''  \citep{a12}. Although recent work has been done to modify the DBSCAN algorithm to address this weakness \citep{crb25}, these approaches still fail to satisfy several criteria necessary for our application, which we describe in the following paragraphs.

Another challenge is the ability to quantify uncertainty on cluster assignment (i.e., uncertainty quantification or UQ). Note that this is not the same as the uncertainty on the parameters describing the clusters (e.g., mean cluster diameter). Both $k$-means algorithms and DBSCAN do not provide uncertainty quantification for cluster membership. \cite{lrpv16} introduced the constrained Dirichlet process model (C-DPM), but does not provide an assessment of UQ, nor does it discuss tuning parameter selection and robustness. Nevertheless, a key strength of the model is that the C-DPM incorporates constrained clustering \citep{wcrs01,wqd14,lrpv16} in a Dirichlet process framework. This constrained clustering offers a natural way to incorporate expert structure.

Constrained clustering restricts the set of possible clusterings, for example, through pairwise constraints between pairs of observations. Pairwise constraints take one of two forms: a must-link (ML) constraint (i.e., two observations must be in the same cluster) or a cannot-link (CL) constraint (i.e., two observations cannot be in the same cluster). Again, neither $k$-means algorithms nor DBSCAN incorporate constrained clustering. We also consider the CRP, which is a distribution over partitions (i.e., clusters) that does not require the specification of the number of clusters, can incorporate any number of features, and can quantify cluster-assignment uncertainty. Multiple extensions include the distance dependent CRP, which has been shown to be a better fit for sequential data \citep{bf11} or other non-exchangeable data \citep{gusb11, mp22}, nested CRP, which allows for distributions on infinitely-branched and infinitely deep trees \citep{bgj10}, spectral CRP, which combines the distance dependent CRP with spectral analysis methods for dimension reduction \citep{smm11, blz23}, or modifying the CRP such that partitions are more probable according to pairwise information \citep{ddt17}. Neither the CRP nor any of its aforementioned extensions incorporate the constraining structure imposed by experts, where it is known $\textit{a priori}$ that observational craters cannot be clustered within an expert. Our work is similar to \cite{wlr20}, who incorporate restrictions on the CRP prior such that clusters must be connected in order to be valid. In our model, however, we modify the CRP to incorporate a CL constraint based on the expert structure, rather than constraining clusterings according to the state of the partition.

Table \ref{tbl:method_summary} compares the methods discussed with respect to the desired properties for a clustering algorithm. No existing approach, to our knowledge, satisfies all of these criteria, motivating the development of the DFCRP.

\begin{table}[!tb]
	\begin{center}
		\caption{Summary of related clustering algorithms and our DFCRP model with respect to the desires and limitations of the crater counter expert data. Check marks indicate whether the method satisfies the corresponding requirement. Only the DFCRP meets all of the necessary criteria.}
		\begin{tabular}{| l | cccccc|}
		\hline
			& k-means & DBSCAN & C-DPM & CRP & CRP Extensions  & DFCRP \\
			\hline
			 Unspecified $k$ & & $\checkmark$ & $\checkmark$  & $\checkmark$  & $\checkmark$ & $\checkmark$  \\
			 Scales w.r.t. features & $\checkmark$ & & $\checkmark$  & $\checkmark$  & $\checkmark$ & $\checkmark$\\
			 UQ on assignment& & & & $\checkmark$  & $\checkmark$ & $\checkmark$  \\ 
			 Incorporates expert structure & & & $\checkmark$ & & & $\checkmark$ \\
			 \hline
		\end{tabular}
		\label{tbl:method_summary}
	\end{center}
\end{table}

The problem of automatically merging unique objects identified by multiple entities (e.g., human experts or computer algorithms) is a problem faced in many disciplines, not just moon crater counting, including metalcasting anomaly detection \citep{dp11}, humans in crowds detection \citep{wn05, lkp24}, traffic light detection \citep{gjxgtc10}, and medical imaging problems such as polyp detection \citep{hc10}, and lung nodule segmentation \citep{cgbpdsp09}. All of these fields take a static image as input and multiple humans and/or computer algorithms detect objects of interest in the image that need to be combined and uncertainty quantified. Recent developments in AI and deep learning methods have changed this object detection landscape \citep{wzh24, zwt26}. Although these methods are able to generate cluster assignments based on features present in the data, they fail to meet all of the criteria outlined above. As an example, such methods generally do not quantify uncertainty on cluster assignments as we require of our model. Thus, the DFCRP provides a solution to the problem of merging identified objects from multiple distinct entities with quantified uncertainty. Before we provide the details of the DFCRP in Section \ref{sec:dfcrp}, we first provide a summary of the CRP in Section \ref{sec:crp}.

%%%%%%%%%%
% CRP SECTION
%%%%%%%%%%
\section{Chinese Restaurant Process}
\label{sec:crp}
The CRP is a probability distribution over partitions of objects. It is conceptually described as follows. Consider a Chinese restaurant with infinitely many tables, where each table can seat infinitely many customers. Let $c_i$ be the table assignment for customer $i$. The first customer enters the restaurant and is placed at the first table (i.e. $c_1 \equiv 1$). Assume that $K$ tables are occupied by the first $n-1$ customers, with $K \leq n-1$. The $n$th customer is assigned to a table based on the following probability rule:
\begin{align}
    \label{eq:crp} % keep reference
   	 Pr_{\crp}(c_n = k \mid c_1, c_2, \ldots, c_{n-1},\alpha) &=
    \begin{cases}
    	\frac{n_k}{n-1+\alpha} &\text{ if } k \leq K \\
    	\frac{\alpha}{n-1+\alpha} &\text{ if } k = K + 1,
    \end{cases}
\end{align}

\noindent where $n_k$ is the number of customers seated at table $k$, $n-1 = \sum_{k=1}^K n_k$ is the number of customers seated in the restaurant, and $\alpha>0$ is a concentration parameter. Larger values of $\alpha$ encourage new customers to sit at unoccupied tables, although the relative influence of $\alpha$ diminishes as $n$ increases. Equation \ref{eq:crp} articulates the ``rich get richer" property, i.e., the more customers seated at a table, the more likely a new customer will sit at that table, which  naturally encourages the number of tables to remain small. For a CRP, a customer represents a data point, a table represents a cluster, and customers seated at the same table constitute data assigned to the same cluster.

The joint distribution on table assignments defined by the CRP leads to a random partition of objects. Consider the joint distribution for table assignments for $n$ customers, $p(c_1, c_2, \ldots, c_n)$. By the chain rule, we can factor the joint distribution as follows:
\begin{align}
    \label{eq:prodrule} % keep reference
    p_{\crp}(\bc) &= p_{\crp}(c_1,c_2,\ldots,c_n) \\ &=Pr_{\crp}(c_2 \mid c_1)\,Pr_{\crp}(c_3 \mid c_1,c_2) \ldots Pr_{\crp}(c_n \mid c_1,c_2,\ldots,c_{n-1}), \nonumber
\end{align}

\noindent where the conditional probabilities are defined by the probability rule in Equation \ref{eq:crp}. The joint distribution of table assignments defined in Equation \ref{eq:prodrule} is referred to as CRP$(\alpha)$.

\iffalse
The number of occupied tables, $K$, is random and chosen by the data.  In practice, the number of possible tables is between 1 and $n$. Although the CRP is a flexible clustering procedure, it does not satisfy all the necessary properties  for automatically merging unique objects identified by multiple entities. Specifically, a cluster cannot have more than one object from each entity assigned to it. In the next section, we introduce the DFCRP, a modification to the CRP, that naturally enforces this constraint. 
\fi

%%%%%%%%%%
% DFCRP SECTION
%%%%%%%%%%
\section{Dysfunctional Family Chinese Restaurant Process}
\label{sec:dfcrp}
The DFCRP is a modification to the CRP described as follows. Like the CRP, consider a Chinese restaurant with infinitely many tables, where each table can seat infinitely many customers. Unlike the CRP, assume customers are members of families and these families are dysfunctional. Family members do not get along with any member of their own family, but do get along with members of other families. The within family dysfunction is so bad, members of the same family cannot sit at the same table, implying no table can seat more than one member of the same family.

The first customer enters the restaurant and is placed at the first table (i.e., $c_1\equiv1$). Let $c_i$ denote the table assignment for the $i$th customer. Let $x_i$ be the family assignment for the $i$th customer, which we treat as known \textit{a priori}.  Let $\bcm$ be the vector of table assignments for all customers except member $i$ from family $x_i$. When the $i$th customer enters the restaurant, assume $K$ tables are occupied. Member $i$ is seated at a table according to the following probability rule:
\begin{align}
    \label{eq:dfcrp}
    Pr^*_{\dfcrp}(c_i = k \mid \bcm, \alpha, \fmly) &=
    \begin{cases}
    0 &\text{ if } n_{k,x_i} > 0 \text{ and } k \leq K \\
    \frac{n_k}{n_{-x_i}+\alpha} &\text{ if } n_{k,x_i} = 0 \text{ and } k \leq K\\
    \frac{\alpha}{n_{-x_i}+\alpha} &\text{ if } k = K + 1,
    \end{cases}
\end{align}
\noindent for $k=1,\ldots,K,K+1$ and $\alpha >0$, where $n_k$ is the number of customers sitting at table $k$, $n_{k,x_i}$ is the number of customers in family $x_i$ sitting at table $k$, and $n_{-x_i} = \sum_{k=1}^K n_k I(n_{k,x_i} = 0)$ is the number of customers seated at tables not occupied by family $x_i$. $I()$ is an indicator function, equal to 1 if the condition is true and 0 otherwise. The asterisk (*) distinguishes this probability rule from the order invariant version, which will be discussed later in Section \ref{subsec:perm}. We note that in Equation \ref{eq:dfcrp} we are conditioning on $\fmly$, however, we do this solely for notational convenience, as we need only condition on $x_1, \ldots, x_i$. 

Conceptually, Equation \ref{eq:dfcrp} says a new customer enters the restaurant, identifies all the occupied tables where a family member is seated, and removes those tables from consideration (i.e., assigns a probability of zero to the tables with the customer's family members). For the remaining tables occupied with only non-family members, the probability of sitting at those tables mirrors that of a CRP($\alpha$). To apply the DFCRP to the identified crater lists, a family represents an expert, a member of family $x_i$ represents an observational crater identified by expert $x_i$, and a table represents a cluster of craters. Not allowing family members to be seated at the same table imposes the constraint that two observational craters from a given expert will not be clustered together---a desired property not feasible with the CRP. 
%Note that Equation \ref{eq:dfcrp} defines a valid probability mass function on $\{1,2,\ldots,K, K+1\}$. 

The joint distribution on table assignments ($c_1,c_2,\ldots,c_n$), defined by the DFCRP, like the CRP, produces a random partition of objects. Consider the joint distribution for table assignments for $n$ customers, $p^*_{\dfcrp}(c_1, c_2, \ldots, c_{n})$. By the chain rule, we can factor the joint distribution as follows:
\begin{align} % keep reference
    \label{eq:prodruledfcrp}
    p^*_{\dfcrp}(\bc \mid \alpha, \fmly) &= p^*_{\dfcrp}(c_1, c_2, \ldots, c_n \mid \alpha, \fmly) \\
    &=\nonumber Pr^*_{\dfcrp}(c_2 \mid c_1, \alpha, \fmly)\,\times \ldots Pr^*_{\dfcrp}(c_n \mid c_1,c_2,\ldots c_{n-1}, \alpha, \fmly),
\end{align}
\noindent where the conditional probabilities are defined by Equation \ref{eq:dfcrp}.

Under the DFCRP, the number of occupied tables, $K$, is still random and determined by the data. However, the DFCRP places a lower bound on the minimum number of tables and an upper bound on table size. Because no two members of the same family can be placed at the same table, the size of the largest family defines the lower bound on the number of occupied tables, where the number of possible tables is between $\max_{k} \sum_{i=1}^n I(x_i = k)$, the size of the largest family, and $n$, the total number of customers. In addition, the dysfunctional family constraint imposes a restriction that the number of customers at a given table cannot exceed the number of families, thus, the potential table sizes are between 1 and the number of families, or the number of unique elements in $\fmly$.  

For clarity, assume there are ten customers ($n=10$) from two families. Four customers are members of the first family and six customers are members of the second family. If the ten customers were seated according to the CRP, the number of possible occupied tables is between 1 and 10; under the DFCRP, the number is between 6 and 10. If the 10 customers were seated according to the CRP, the possible table sizes are between 1 and 10; for the DFCRP, the possible table sizes are only 1 or 2. In general, the DFCRP tends to form more, smaller tables relative to the CRP. 

\subsection{Permutation Invariance}
\label{subsec:perm}
One difference between the the standard CRP and the DFCRP is that the standard CRP is an exchangeable distribution. A consequence of exchangability is that the order by which items are assigned to clusters within a partition does not change the probability of said partition occurring. In the DFCRP however, this assumption is violated, owing to the nature of the dysfunctional families. This order invariant nature is an important attribute of the DFCRP because we want to ensure that the probability of a partition is constant regardless of the permutation of the items. For example, modeling the number of clusters (i.e. craters on the moon) should result in probabilities of partitions that are invariant with respect to the permutation of the data (i.e. experts' identified crater observations). The unmodified DFCRP assigns different probabilities to the same partition, depending on the order in which the items are allocated to the partition. Probabilities and distributions under this unmodified DFCRP are denoted above with an asterisk (*).   

To restore order invariance, we introduce a permutation parameter.  We place a prior on the permutation parameter and then marginalize it out of the model.  This approach, inspired by \cite{ddt17}, allows us to assign a single probability to each partition regardless of the permutation.  This ensures order invariance on this modified DFCRP, but does not go as far as achieving exchangeability. The permutation parameter, denoted $\bsigma$, relates to the sequential order in which data are assigned to clusters. To demonstrate, consider an example with customers A, B, and C. One permutation, denoted $\bsigma=(C, A, B)$ indicates that the first customer to be allocated is C, then A, and finally B. The permutation parameter $\bsigma$ can therefore represent any of the $n!$ permutations of the $n$ customers/items.

Revising Equation \ref{eq:dfcrp} to include a permutation parameter we obtain the following:
\begin{align}
    \label{eq:dfcrpFinal}
    Pr_{\dfcrp}(c_{\sigma_{i}} = k \mid \bcmperm{i-1}, \alpha, \fmly, \bsigma) &=
    \begin{cases}
    0 &\text{ if } n_{k,x_{\sigma_{i}}} > 0 \text{ and } k \leq K \\
    \frac{n_k}{n_{-x_{\sigma_{i}}}+\alpha} &\text{ if } n_{k,x_{\sigma_{i}}} = 0 \text{ and } k \leq K\\
    \frac{\alpha}{n_{-x_{\sigma_{i}}}+\alpha} &\text{ if } k = K + 1,
    \end{cases}
\end{align}
Where $\sigma_{i}$ represents the $i$th item allocated, thus $c_{\sigma_{i}}$ is the cluster label of the $i$th item allocated.  Thus $\bsigma$ defines a specific order in which the items are allocated.  Additionally, $x_{\sigma_{i}}$ denotes the family of the $i$th item allocated and $n_{-x_{\sigma_{i}}}$ is the number of items allocated (thus far) which are not in family $x_{\sigma_{i}}$. 

As originally defined, the joint distribution on table assignments under the DFCRP in Equation \ref{eq:prodruledfcrp} implicitly assumes a fixed permutation $\bsigma$. Modifying this equation to explicitly state an order of allocation we get:
\begin{equation}
  \label{eq:dfcrppmfsig}
  p_{\dfcrp}(\bc \mid \alpha, \fmly, \bsigma) = 
  \prod_{i=2}^{n} Pr_{\dfcrp}(c_{\sigma_{i}} \mid \bcmperm{i-1}, \alpha, \fmly, \bsigma).
\end{equation}

Generally, the permutation parameter $\bsigma$ can be seen as a nuisance parameter because, in most contexts, probabilities of partitions should be equal regardless of the order in which the items are allocated. To represent this lack of prior knowledge for $\bsigma$, we place a uniform prior on $\bsigma$, such that $p(\bsigma)=(n!)^{-1}$, as suggested in \cite{ddt17}. 

Thus we can represent the marginal distribution of a partition by summing out all possible permutations $\bsigma$, as follows: 
\begin{align}
    \label{eq:perm} % keep reference
  p_{\dfcrp}(\bc \mid \alpha, \fmly) =\sum_{\bsigma\in \mathcal{S}}p_{\dfcrp} (\bc \mid \alpha, \fmly, \bsigma) \, p(\bsigma) =\frac{\sum_{\bsigma\in \mathcal{S}}  p_{\dfcrp}(\bc\mid \alpha, \fmly, \bsigma)}{n!}  
\end{align}
Where $\mathcal{S}$ denotes the set of all possible permutations of $n$ elements, and $p_{\dfcrp}(\bc\mid \alpha, \fmly, \bsigma)$ denotes the distribution of the partition $\bc$ given a permutation $\bsigma$ (as defined in Equation \ref{eq:dfcrppmfsig}). It is not computationally feasible to enumerate all possible $\bsigma$ for large $n$, thus in practice, we marginalize over $\bsigma$ using Monte Carlo sampling \textit{a priori}, and using MCMC \textit{a posteriori}. See Section B in the supplementary material for further details.
 
When referring to $p_{\dfcrp}(\bc \mid \alpha, \fmly)$ we have marginalized over the permutation parameter as follows:
\begin{align}
    \label{eq:prodruleddfcrpw/perm} % keep reference
  p_{\dfcrp}(\bc \mid \alpha, \fmly) =\sum_{\bsigma\in \mathcal{S}} p_{\dfcrp} (\bc \mid \alpha, \fmly, \bsigma) \, p(\bsigma).
\end{align}
Here, $p_{\dfcrp}(\bc \mid \alpha, \fmly)$ produces a partition distribution which is invariant to the order in which items are allocated.  We denote random partitions with the DFCRP as defined in Equation \ref{eq:prodruleddfcrpw/perm} as:
$$\bc \mid \alpha, \fmly \sim DFCRP(\alpha, \fmly).
$$
We assume that the permutation parameter $\bsigma$ has already been marginalized from the model unless otherwise specified. 

%%%%%%%%%%
% PERMUTATION MODIFICATION
%%%%%%%%%%
\subsection{Modifying the Permutation Proposal Algorithm}
\label{subsec:permutationprop}

The nonexchangeable DFCRP prior is substantially more expensive computationally than the CRP prior, motivating the need for a more efficient sampling scheme. One possibility is to exploit the partial exchangeability present in the distribution. Under full exchangability, the DFCRP prior in Equation \ref{eq:dfcrppmfsig} would be equal for all values of $\bsigma$. As we have discussed, this is not the case. However, there do exist some permutations $\bsigma_1$ and $\bsigma_2$ where the DFCRP prior is equal under both permutations. By strategically swapping items in the permutation such that the original and proposed permutations both produce an equivalent prior evaluation under the DFCRP, we can explore the posterior of the permutation parameter without needing to recompute the nonexchangable DFCRP prior. Finding such permutations, however, is nontrivial, and we were unable to identify a general procedure for determining when two permutations yield equal prior evaluations under the DFCRP.

Another alternative is to implement a proposal algorithm that will only reallocate the last item in the permutation. Although a single update of the partition requires computing the full, nonexchangable DFCRP prior, the update of the last item is computationally equivalent to the CRP because we condition on the allocations of the items before it (as in Equation \ref{eq:dfcrpFinal}). This means we can use a CRP-style update for the last item in a partition. Thus, by editing our permutation proposal mechanism to cycle through the items of a permutation in a specific way, we can ensure the entire partition is updated, albeit with less efficiency than before. Accordingly, we modified our permutation proposal algorithm to randomly swap an item in the permutation with the last item. We refer to the Gibbs sampler with this altered permutation Metropolis sampler as the \emph{permutation-modified Gibbs sampler}.  

The reduced efficiency imposed by this modification necessitates redefining our understanding of an "iteration". Previously, one iteration corresponded to an update of the entire partition according to the non-exchangable DFCRP prior. With this modification, however, the proportion of the partition that is updated each iteration is only $n^{-1}$, where $n$ is the number of customers in the dataset. Thus, we treat $n$ iterations as a complete "scan" of the partition. Although not all $n$ items of the partition will be updated in $n$ iterations, we consider these scans as a sufficient step in the posterior space of the partition. For the rest of the paper, we use the term "scan" to mean $n$ iterations of the sampler. Running the permutation-modified Gibbs sampler on the full dataset yielded an average permutation acceptance rate of $33.5\%$. This acceptance rate per iteration suggests that over a scan, a substantial fraction of partition allocations are updated, ensuring adequate exploration of the posterior. See Section B.3 in the supplementary material for further details.

%%%%%%%%%%
%%% DFCRP Mixture Model subSection 
%%%%%%%%%%
\section{DFCRP Mixture Model}
\label{sec:dfcrpmm}
We consider the DFCRP mixture model that jointly models the data ($\by$),  the cluster-specific parameters ($\btheta$), the datum-specific cluster assignments ($\bc$), and the DFCRP concentration parameter ($\alpha$), conditional on the family assignments ($\fmly$). We model these components hierarchically as follows:
\begin{align} % keep reference
    \label{eq:joint}
    p(\by,\btheta,\bc,\alpha \mid \fmly) = p(\by \mid \btheta,\bc) \, p(\btheta) \, p_{\dfcrp}(\bc \mid \alpha, \fmly) \, p(\alpha),
\end{align}
notationally suppressing some hyperparameters.
The joint distribution on datum-specific cluster assignments is the DFCRP($\alpha$, $\fmly$):
\begin{align*}
    \bc \mid \alpha, \fmly &\sim \text{DFCRP}(\alpha, \fmly),
\end{align*}
again where $\bc$ is a vector of cluster label assignments. Given a cluster assignment $\bc$, the number of clusters $K$ is known and is the number of unique cluster labels in $\bc$.  

The data, $\by$, conditioned on the cluster assignments, $\bc$, and the table parameters, $\btheta$, are assumed to be conditionally independent from a multivariate Normal (MVN) distribution:
\begin{align*}
    p(\by \mid \btheta,\bc) &= \prod_{i=1}^{n} \text{MVN}(\by_i \mid \bmu_{c_i},\bSigma_{c_i}).
\end{align*}

Although we use the MVN distribution in our application, the cluster-specific likelihood and associated priors can be tailored to the needs of a given application. Note that there are $K$ cluster-specific means and covariances for these MVN distributions. Also note the dimension of the MVN distribution is determined by the number of features on which to cluster, which we will assume to be $p$.

The prior $p(\btheta)$ encompasses the prior on both the cluster-specific means and covariances. The cluster-specific parameter $\bSigma_k$ represents the covariance matrix associated with a specific cluster $k$. For our application, we define the prior $p(\bSigma)$ via priors on specific elements of the $\bSigma$ matrix, rather than on the matrix as a whole. We define a general cluster-specific covariance matrix ($\bSigma$), with columns $(X, Y, l\text{D})$, like so:

\begin{align*}
\renewcommand{\arraystretch}{1.2}
\bSigma =
\begin{bmatrix}
\sigma^2_{x} & 0 & \sigma^2_{xd} \\
0 & \sigma^2_{x} & \sigma^2_{xd} \\
\sigma^2_{xd} & \sigma^2_{xd} & \sigma^2_{d}
\end{bmatrix}.
\end{align*}

\noindent Note that for notational convenience, the subscript $x$ is used for both spatial coordinates. We assign the following priors to the diagonal elements of this matrix. 

\begin{align}
\label{eq:varpriors}
\sigma^2_{x} \mid l\text{D} &\sim \text{Gamma}\!\left(\tau_{x}\kappa_{x}\,l\text{D}^{\eta_{x}},\; \tau_{x}\right), \qquad
\sigma^2_{d} \mid l\text{D} \sim \text{Gamma}\!\left(\tau_{d}\kappa_{d}\,l\text{D}^{\eta_{d}},\; \tau_{d}\right),
\end{align}

\noindent where $\kappa$ and $\eta$ represent the power law coefficient and exponent respectively, $\tau_{x}$ denotes the precision of the prior on $\sigma^2_{x}$, and $\tau_{d}$ denotes the precision of the prior on $\sigma^2_d$. We believe \textit{a priori} that these variances should be dependent on the $l$D. Visually, how similar craters must be in terms of their features will depend on how large those craters are. Accordingly, we modeled the relationship between the variance and the $l$D via a power law relationship. We then made this relationship the mean for our prior distributions, as enumerated above.

The off-diagonals of the covariance matrix $\bSigma_k$ represent the covariance between the features of the data. We set the covariance between the spatial coordinates to be 0. We also consider the covariance between the spatial coordinates and the diameter of the crater to be at or near 0, and accordingly we assign those covariances the following prior.

\begin{align}
\label{eq:covarprior}
\lambda \sim 2 \;\text{Beta}(a_{\lambda}, b_{\lambda}) - 1, 
\qquad \sigma^2_{xd} = \lambda \sqrt{\frac{\sigma^2_{x}\sigma^2_d}{2}}.
\end{align}

\noindent In order for a covariance matrix to be positive semi-definite, we require that $\sigma^2_{xd} + \sigma^2_{xd} \leq \sigma^2_{x}\sigma^2_d$, which follows from the requirement that all principal minors of a covariance matrix be nonnegative. The $\sqrt{\sigma^2_{x}\sigma^2_d/2}$ term ensures that this condition is never violated.

The other component of $p(\btheta)$ involves the prior on the cluster-specific means. We employ a MVN prior distribution to ensure conditional conjugacy with the posterior. Let

\begin{align}
\label{eq:muprior}
\bmu_k \sim \mathcal{N}(\bmu_0, \bSigma_0).
\end{align}

\noindent We provide rationale for our selections of $\bmu_0$ and $\bSigma_0$, along with the other hyperparameters in this section, in the following subsection.

The hyperparameter $\alpha$ influences the number of clusters. Smaller values of $\alpha$ tend to lead to fewer clusters, while larger values of $\alpha$ correspond to more clusters. To add more flexibility to our DFCRP model, we use a Gamma prior distribution with parameters $a_{\alpha}$ and $b_{\alpha}$. That is, $p(\alpha)\sim \text{Gamma}(a_{\alpha},b_{\alpha})$. Details on the sampling procedure from the conditional distribution of $\alpha$ are given in Section A of the supplementary material. While this prior introduces greater flexibility, the DFCRP model itself already places a constraint on the number and size of clusters independent of the choice of $\alpha$, with an upper bound on cluster size equal to the number of families and a lower bound on the number of possible clusters equal to the size of the largest family. In the context of the DFCRP, $\alpha$ can be interpreted as the unnormalized probability of new clusters forming above this lower bound on cluster counts.

\subsection{Hyperparameter Selection}
\label{subsec:hyper}
There are several hyperparameters in the DFCRP mixture model that must be specified. Naturally, hyperparameter specification should be specific to the application. In what follows, we provide a basis for our specification of hyperparameters specific to our application of clustering craters on the moon. For clarity, all values associated with the dataset presented in this section are expressed in units of pixels. The data are recorded on a coordinate grid where $x$ ranges from $64$ px to $4067$ px, $y$ ranges from $-2221$ px to $10$ px, and the log diameter ($l$D) of the observed craters ranges from $2.89$ to $6.23$.

As seen in Equation \ref{eq:varpriors}, the values of $\kappa_{x}$, $\kappa_d$, $\eta_{x}$, and $\eta_d$ quantify the relationship between the within-cluster variability of crater features and the average $l$D of the craters present in the cluster. To determine these values, we generated craters of varying $l$Ds to ascertain a reasonable level of tolerance for the three features in our application. We concluded that specifying $\kappa_{x} = 0.08$, $\kappa_d = 0.124$, $\eta_{x} = 4.5$, and $\eta_d = -0.8$ quantified an accurate relationship between $l$D and within-cluster feature variance. We also selected $\tau_{x} = 1$ and $\tau_d = 100$, which reflects a greater confidence in the hyperparameter selection for $\sigma^2_d$ as compared to that of $\sigma^2_{x}$.

In Equation \ref{eq:covarprior}, the choice of $a_\lambda$ and $b_\lambda$ reflect the $\textit{a priori}$ belief about how the $x$ and $y$ coordinates of craters in a cluster are associated with the $l$D of those craters. They control a centered and scaled beta distribution on $\lambda$, which is a ratio of the covariance to its defined maximum, $\sqrt{\sigma^2_{x}\sigma^2_d/2}$. Since our intuition $\textit{a priori}$ is that the spatial coordinates of craters within a cluster would have little to no relationship with the diameter of those same craters, we set $a_\lambda = 100$ and $b_\lambda = 100$ for a tight distribution on $\lambda$ centered at zero.

It is important to note that these specifications are permissive enough to allow reasonably similar craters to be clustered together, yet restrictive enough to discourage unintuitive clustering. The conditional distributions on $\sigma^2_{x}$, $\sigma^2_{d}$, and $\sigma^2_{xd}$ keep cluster-specific covariance matrices small, which reduces the likelihood of clustering dissimilar craters. 
%and additionally reduces the likelihood of substantially different posterior draws of the partition. 
With these strictly specified priors, we expect fairly low variability in the posterior of the partition $\bc$.
%We thus expect relatively low variability in our posterior draws of cluster assignments except in cases where the data favors several alternative clusterings.

Additionally, we need to specify the values of $\bSigma_0$ and $\bmu_0$ in Equation \ref{eq:muprior}; these values should allow for sufficient flexibility in the model. These parameters represent where cluster means may be located in an image, and their specification is accordingly data driven. To specify $\bmu_0$ for use on the full dataset, we used the midpoint of the data, $\bmu_0 = (2068.2, -1105.5, 3.8)$, Via empirical testing, we also found that $\bSigma_0 = \operatorname{diag}(920^2, 600^2, 0.65^2)$ served as a reasonable value given the nature of the full dataset.

Finally, we selected the hyperparameters for the gamma prior on $\alpha$ based on our beliefs $\textit{a priori}$ about the number of craters above the minimum, the size of the largest family. Our lack of intuition on the matter led us to place a very diffuse prior on $\alpha$: we set $a_{\alpha} = 3$ and $b_{\alpha} = 0.04$. This prior contributes to more variability in the posterior.

For details on the permutation hyperparameter $\bsigma$, see Section \ref{subsec:perm}.

%%%%%%%%%%
%%% Posterior Distribution subSection
%%%%%%%%%%
\section{Posterior Considerations}
\label{sec:posteriorsampling}

The posterior of particular interest is that of the partition, conditional on the data, family assignments and concentration parameter $\alpha$. To sample from $p(\bc \mid \by, \alpha, \fmly)$, we implement a Gibbs sampler (\cite{gg84}) to alternately update $\bc$, $\alpha$, $\bsigma$, and the cluster-specific parameters $\btheta$.

After sampling from $\bc$ $n$ times (i.e. after one scan), we update $\bSigma$ via a Metropolis sampler. We utilize a MVN proposal distribution on each of the elements of the $\bSigma$ matrix, like so:

$$(\sigma^{2 \; (m)}_{x}, \sigma^{2 \; (m)}_{d}, \sigma^{2 \; (m)}_{xd}) \sim \text{MVN}(\mu = (\sigma^{2 \; (m-1)}_{x}, \sigma^{2 \; (m-1)}_{d}, \sigma^{2 \; (m-1)}_{xd}), \; \bSigma = \boldsymbol{S}),$$
where $\boldsymbol{S}$ denotes the proposal matrix for the components of $\bSigma$. Although the selection of $\boldsymbol{S}$ will need to be tuned depending on the application, we set $\boldsymbol{S} = \text{diag}(3000, 0.9, 0.2)$. 
%This specification of $\boldsymbol{S}$ produced relatively low acceptance rates, but extensive tuning did not lead to substantial improvements. This behavior is likely due to the informativeness of the prior $p(\bSigma)$, which concentrates posterior mass in a relatively small region of the parameter space. 
After sampling from the posterior of $\bSigma$, we can readily sample from the posterior of $\bmu$ given the data and the current value of $\bSigma$, as follows:
\begin{align}
\begin{split}
    \bmu_k &\mid \by_k, \; \bSigma_k \sim \mathcal{N}(\bmu_k^*, \bSigma_k^*), \ \ \,
\text{where}\\
\bSigma_k^* &= \left( \bSigma_0^{-1} + n_k \bSigma_k^{-1} \right)^{-1},\ \ 
\text{and}\\
\mu_k^* &= \bSigma_k^* \left( \bSigma_0^{-1} \bmu_0 + n_k \bSigma_k^{-1} \boldsymbol{\bar{y}}_k \right).
\end{split}
\end{align}
\noindent Here, $\by_k$ represents the data assigned to cluster $k$, and $\boldsymbol{\bar{y}}_k$ represents the mean of that data.

Informally, a draw from the full conditional distribution $p(\bc \mid \by, \alpha, \fmly)$ requires answering two questions. The first is, ``What is the probability datum $\byperm$ will be assigned to cluster $k$ according to the DFCRP($\alpha$, $\fmly$)?'' where $\byperm$ is more likely to be assigned to larger clusters than to smaller ones. The second is, ``How well does $\byperm$ `fit in' with the other data points currently assigned to cluster $k$?'' If multiple clusters are of equal size, $\byperm$ is more likely to be assigned to the cluster with data ``most similar'' to $\byperm$. If $\byperm$ ``fits in'' with two clusters equally well, it is more likely to be assigned to the larger cluster. Similar to Neal's Algorithm 8, we obtain $M$ scans (see Section \ref{subsec:permutationprop}) from the permutation-modified Gibbs sampler as outlined in Algorithm \ref{alg:gibbs}. 
\spacingset{1} 
\RestyleAlgo{boxruled}
\LinesNumbered
\begin{algorithm}[ht]
  \caption{Permutation-Modified Gibbs Sampler\label{alg:gibbs}}
  \begin{algorithmic}[1]
  \STATE Convert the diameter feature of the data to the log scale
  \STATE Initialize: Initialize the sampler by assigning each crater to its own cluster. Call this $\bc^{(0)}$
  \STATE Assign each cluster-specific mean $\bmu$ to be the feature data for the crater assigned to that cluster. Draw \\ each cluster-specific covariance matrix from the prior $p(\bSigma)$
  \FOR{$m$ in $1,2,\ldots,M$}
    \STATE Draw $\alpha^{(m)}$ using the Metropolis-Hastings sampler (see Section A in the \\supplementary material)
    \FOR{$i$ in $1,2,\ldots,n$} 
        \STATE Draw $\bsigma^{(i)}$ using the Metropolis-Hastings sampler (see Section B.2 in the \\supplementary material)
        \STATE Remove datum $\by_{\sigma_n}$ from cluster $c_{\sigma_n}$
        \STATE Define $K$ as the set of clusters in $\bc^{(m-1)}$ with $\by_{\sigma_n}$ removed
        \STATE Set $\bm{k}_{-x_{\sigma_n}}$ to be a vector of all clusters in $K$ without a member of family $x_{\sigma_n}$ \\present
        \STATE Initialize $\bm{z}$ as an empty vector with length $|\bm{k}_{-x_{\sigma_n}}|+1$ 
        \label{alg:Cset}
        \FOR{$j$ in $\bm{k}_{-x_{\sigma_n}}$} \label{alg:startK}
            \STATE Compute $p(\by_{\sigma_n} \mid \by_{\sigma_{-n}}, \bc, \btheta) \cdot Pr_{\crp}(c_{\sigma_n} = j \mid c_{\sigma_1}, c_{\sigma_2} \ldots c_{\sigma_{n-1}}, \alpha^{(m)}) := z_j$
        \ENDFOR \label{alg:endK}
        \STATE Compute $p(\by_{\sigma_n} \mid \by_{\sigma_{-n}}, \bc, \btheta) \cdot Pr_{\crp}(c_{\sigma_n} \mid c_{\sigma_1}, c_{\sigma_2} \ldots c_{\sigma_{n-1}}, \alpha^{(m)})$ for $c_{\sigma_n}$ equal \\ to a new cluster and store it as the final entry in the vector $\bm{z}$
      \STATE Normalize $\bm{z}$ to obtain a probability vector $\tilde{\bm{z}}$, then sample $c^{(i)}_{\sigma_n} \sim \text{Categorical}(\tilde{\bm{z}})$
      \STATE Set $\bc^{(i)} = (c^{(i-1)}_{\sigma_1},c^{(i-1)}_{\sigma_2},\ldots,c^{(i)}_{\sigma_n})'$. $\bc^{(i)}$ constitutes a joint draw of $\bc$ \\from $p(\bc \mid \by, \alpha, \fmly)$
      \ENDFOR
      \STATE Update $\btheta$ as detailed in Section $\ref{sec:posteriorsampling}$
  \ENDFOR
  \end{algorithmic}
\end{algorithm}
\spacingset{1.8} 

%%%%%%%%%%
% NEIGHBORHOOD MODIFICATION
%%%%%%%%%%
\subsection{Neighborhood-Modified Gibbs Sampler}
\label{subsec:neighborhood}
The Gibbs sampler of Algorithm \ref{alg:gibbs} suggests a computational speed-up. Consider Figure \ref{fig:data}. Craters that are clustered together will be spatially proximate. For each data point $\byperm$, however, the Gibbs sampler requires evaluating the predictive likelihood for every cluster that does not contain a member of family $x_{\sigma_i}$. That is, for each iteration of the Gibbs sampler, the sampler checks every datum's fit with every possible cluster (lines \ref{alg:startK} through \ref{alg:endK} of Algorithm \ref{alg:gibbs}). Due to the large number of possible clusters and proximate nature of those observational craters clustered together, we expect many of the predictive likelihood evaluations to be effectively zero. To account for this redundancy, we introduce a quantity, denoted $\rho$, that defines a neighborhood for each datum. For our application, the neighbors of datum $\byperm$ are all data points $\by_{\sigma_i'}$ such that $\sqrt{(X_{\sigma_i} - X_{\sigma_i'})^2 + (Y_{\sigma_i} - Y_{\sigma_i'})^2} \leq \rho$, where $X_{\sigma_i}$ and $Y_{\sigma_i}$ are the longitude and latitude  for data point $\byperm$. More generally, a neighborhood can be defined with any distance metric over features. When sampling the cluster assignment for $\byperm$, rather than evaluate the predictive likelihood for all possible clusters under the DFCRP, we propose only doing so for clusters with at least one neighbor of $\byperm$ assigned to it. That is, we propose changing line \ref{alg:Cset} of Algorithm \ref{alg:gibbs} to ``Set $\bm{k}_{-x_{\sigma_i}}$ to be a vector of all clusters in $K$ without a member of family $x_{\sigma_i}$ present and at least one neighbor of $\byperm$ is assigned." We refer to the Gibbs sampler of Algorithm \ref{alg:gibbs} with line \ref{alg:Cset} changed as the \emph{neighborhood-modified Gibbs sampler}.

Running this neighborhood-modified Gibbs sampler requires selecting $\rho$. Small values of  $\rho$ lead to decreases in computation time by reducing the number of predictive likelihood evaluations but risk changing the posterior distribution. Our goal is to select the smallest $\rho$ such that the posterior distributions for quantities of interest estimated from the neighborhood-modified Gibbs sampler are consistent with the estimates from the original sampler. 

\begin{figure}[!tb]
    \centering
    \includegraphics[width=284.2pt]{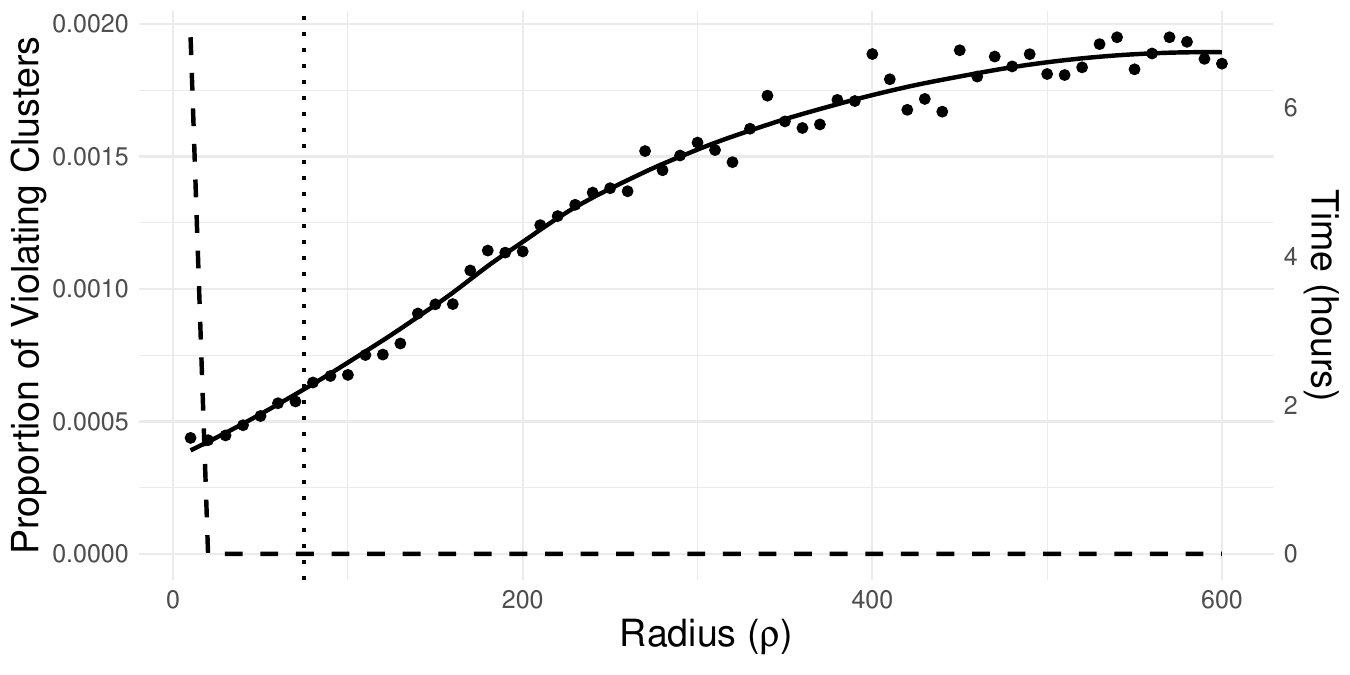}
    \caption{The average proportion of violating clusters (clusters containing items separated by distances exceeding the radius $\rho$) within a subset of crater data as a function of radius $\rho$ (dashed line), plotted alongside the computation time required to obtain 10,000 scans (solid line). The violating-cluster proportion is zero at $\rho = 30$ px; however, to account for structure outside the subset, we adopt an expanded radius of $75$ px (dotted line) on the full dataset. Computation time levels off for $\rho \geq 600$ px.}
    \label{fig:grplot}
\end{figure}

To select such a $\rho$, we first ran 200 chains of the permutation-modified Gibbs sampler—without any radius restriction present—on a subset of the crater data (see Section D of the supplementary material for details on the reduced dataset and the parameters specified thereon). After discarding burn-in and thinning, we utilized 160,000 posterior draws of the partition to evaluate the radius constraint. For this subset, $\text{max}\Big(\sqrt{(X_{\sigma_i} - X_{\sigma_i'})^2 + (Y_{\sigma_i} - Y_{\sigma_i'})^2}\Big) \approx 728 \; \text{px}$. Thus, running the neighborhood-modified Gibbs sampler with values of $\rho \geq 728$ is equivalent to running the sampler without the neighborhood modification. As we consider posterior draws of the partition with decreasing values of $\rho$, more of the clusters in the partition violate the constraint. At the extreme, a value of $\rho = 0$ will consider each cluster of size $> 1$ a violation of the constraint. Thus, we considered each of the draws under different values of $\rho$ to determine the proportion of clusters in the partition that violate the constraint. At a radius of 30 pixels, none of the draws had clusters that violated the radius constraint. Although this radius did not noticeably change the posterior draws obtained from this subset of the data, large identified craters present in the full dataset were more spatially distant yet clearly referred to the same crater structure. The most distant of these large craters was about 70 pixels apart, and accordingly we used a radius of 75 pixels to greatly reduce computation time but allow for reasonable clustering (see Figure \ref{fig:grplot}).

Computation run time improves when  $\rho < 600$, as can be seen in Figure \ref{fig:grplot}. The run time for $\rho=70$ is roughly 2 hours, while the run time for $\rho \geq 600$ is just over 6.5 hours for the selected subimage. Based on these results, we conclude that running the neighborhood-modified and permutation-modified Gibbs sampler for $\rho \approx 75$ generates posterior draws that are consistent with those produced by running the permutation-modified Gibbs sampler at a significant computational speedup.

%%%%%%%%%%
% SIMULATION STUDY
%%%%%%%%%%
\section{Simulation Study}
\label{sec:sim_study}
% \cite{lrpv16} concluded that the benefits of CL constraints to clustering are unclear and may even make clustering performance worse. 
We perform a simulation study comparing the DFCRP to the CRP. The purpose of the simulation study is to investigate what benefits, if any, the DFCRP provides over the CRP when family structure is available and incorporated into the model. Conceptually, one would expect the DFCRP to do no worse, on balance, than the CRP. This is because the DFCRP constrains the space of possible partitions relative to the CRP, where the true partition belongs to the reduced partition space. Thus, the DFCRP has correct and potentially useful information at its disposal unavailable to the CRP. Data with three features ($X_k, Y_k, D_k$) and family structure was simulated according to the procedure outlined in Section C of the supplementary material.
For both the DFCRP and the neighborhood-modified DFCRP, the permutation-modified Gibbs sampler was run for 10000 scans, throwing out the first 2000 scans as burn-in. The CRP was run for 10000 iterations, throwing out 2000 iterations as burn-in. The hyperparameters for the DFCRP and CRP were selected as suggested in Section D of the supplementary material for the reduced dataset and were kept consistent between models. Thus, the only difference between the DFCRP and the CRP fit for a given dataset was the model itself, not the hyperparameters. 

For each dataset and model, we computed the adjusted Rand index \citep{r71}, a similarity metric between two different partitions of a common collection of objects.  The index compared the partitions from the fitted model (DFCRP or CRP) and the true partition. The adjusted Rand index is between -1 and 1, where 1 represents identical partitions, and 0 represents partitions clustered similarly by chance. For this application, a large adjusted Rand index indicates the sampled posterior partition is similar to the truth. Results are shown in Table \ref{tab:randindex}. 

The DFCRP greatly outperformed the CRP with a superior adjusted Rand index for 493 out of the 500 simulations. This result is expected as the DFCRP uses additional information that the CRP does not incorporate. Namely, the DFCRP is constrained to sample partitions where no cluster has more than one family member. The CRP has no such constraint. On average, 5.4$\%$ of all clusters identified by the CRP contained multiple members of the same family, where no true clustering has more than one family member per cluster. Thus, the DFCRP samples partitions over a constrained space, where the true partition is a member of the constrained space. 

\begin{table}[!tb]
\centering
\caption{Posterior mean adjusted Rand index (ARI) summaries for DFCRP and CRP fits across 500 simulated datasets. The ARI measures similarity between the partition inferred by the model and the true partition. Columns display summary statistics of the 500 posterior mean ARI values. Each row corresponds to a model applied to the simulated data, and the final row reports paired posterior mean differences between the DFCRP (without the radius constraint) and the CRP. The DFCRP achieved a higher ARI than the CRP in 493 of 500 simulations.}
\label{tab:randindex}
\begin{tabular}{rccccc}
  \hline
 \textbf{Type} & \textbf{Minimum} & \textbf{25th Percentile} & \textbf{Mean} & \textbf{75th Percentile} & \textbf{Maximum} \\ 
  \hline
  DFCRP & 0.890 & 0.988 & 0.989 & 0.999 & $\sim$ 1 \\ 
  DFCRP w/Radius & 0.884 & 0.986 & 0.989 & 0.999 & $\sim$ 1 \\
  CRP & 0.736 & 0.944 & 0.958 & 0.985 & 0.999 \\ 
  Difference & -0.005 & 0.010 & 0.031 & 0.047 & 0.176 \\ 
   \hline
\end{tabular}
\end{table}

In addition, we used the same 500 datasets to evaluate another drawback to using a model modified from the original CRP. \citet{mh14} found that Pitman-Yor Process Mixtures, including the CRP, overestimate the number of clusters in a partition on average. In the context of our application, the DFCRP model is also susceptible to overestimate the number of true craters for a given dataset. Recall that in their analysis, R14 focused on clusters of size 5 or larger, suggesting that if 5 or more experts independently identified an object, that object is likely a true crater, and referred to these true craters as consensus clusters. To correct for this effect, we propose selecting a consensus cluster count as a more accurate measure for the number of clusters where the true cluster counts is unknown. Table \ref{tab:clustercounts} shows the posterior average number of clusters for both the DFCRP and the CRP.

When considering only the clusters of size 1 or larger, both the DFCRP and the CRP overestimate the true number of clusters, which is 30 for this simulation study. This is consistent with the findings of \citet{mh14}. The DFCRP model most accurately predicts the true number of clusters when we examine clusters of size 3 to 4, but the error is minimal even for clusters of sizes 2 and 5. Accordingly, we conclude that when estimating the true number of clusters for an unknown dataset, mid-range consensus clusters provide a reasonable value. Finally, we note that the number of clusters for clusters with a smaller size is greater with the DFCRP model; this is due to the family constraints imposed by the DFCRP. These restrictions inflate cluster counts beyond the already inflated cluster counts from the CRP, and therefore using consensus craters as estimates is all the more necessary. As cluster size increases, however, the effect of the likelihood dominates, and the partition in these instances is primarily data-driven.

\begin{table}[!tb]
\centering
\caption{Average posterior cluster counts for the DFCRP, DFCRP with radius, and CRP fits across 500 datasets. Each column corresponds to a different minimum size threshold used to define a true cluster. The true number of clusters in each dataset is 30.}
\label{tab:clustercounts}
\begin{tabular}{rcccccc}
  \hline
 \textbf{Type} & \textbf{Size 1} & \textbf{Size 2} & \textbf{Size 3} & \textbf{Size 4} & \textbf{Size 5} & \textbf{Size 6} \\ 
  \hline
   DFCRP & 42.411 & 30.500 & 30.009 & 29.831 & 28.056 & 19.194 \\ 
   DFCRP w/Radius & 42.400 & 30.500 & 30.009 & 29.833 & 28.063 & 19.196 \\
   CRP & 40.473 & 29.663 & 29.256 & 29.091 & 27.441 & 19.246 \\  
   \hline
\end{tabular}
\end{table}

%%%%%%%%%%
% MOON CRATER IMAGE ANALYSIS 
%%%%%%%%%%
\section{Analysis of the Lunar Highlands Image}
\label{sec:full}

We ran 200 chains of the neighborhood-modified and permutation-modified Gibbs sampler, with $\rho=75$, on the data from the full lunar highlands image (Figure \ref{fig:data}) for 10,000 scans each, throwing out the first 5,000 scans as burn-in. We also based our posterior analysis on a thinned sample of the chains, selecting every 10th scan, resulting in 100,000 samples. Hyperparameters were selected as outlined in Section \ref{subsec:hyper}. The DBSCAN estimates (from R14) and the DFCRP posterior means with 95\% credible intervals are presented in Table \ref{tab:fullimagesummary}.

\begin{table}[!tb]
\centering
\caption{DBSCAN estimates and DFCRP posterior means, along with 95\% credible intervals from the posterior draws of the DFCRP. The rows denote the size of the crater (small, medium, large). Small craters have diameters of 18–50 px, medium craters 50–100 px, and large craters exceed 100 px. The row names also denote the minimum size required to consider the cluster a consensus cluster (4, 5, or 6 craters).}
\label{tab:fullimagesummary}
\begin{tabular}{rccc}
  \hline
 & \textbf{DBSCAN Est} & \textbf{DFCRP Mean} & \textbf{DFCRP 95\% Credible Interval} \\ 
  \hline
  Small Size $\geq$ 4 Clusters  & 795 & 809.62 & (803, 816)  \\ 
  Small Size $\geq$ 5 Clusters  & 754 & 742.81 & (737, 749)  \\ 
  Small Size $\geq$ 6 Clusters  & 714 & 684.27 & (679, 690)  \\ 
  \hline
  Medium Size $\geq$ 4 Clusters  & 101 & 98.98 & (92, 106)  \\ 
  Medium Size $\geq$ 5 Clusters  & 94 & 92.41 & (86, 99)  \\ 
  Medium Size $\geq$ 6 Clusters  & 87 & 85.46 & (80, 91)  \\ 
  \hline  
  Large Size $\geq$ 4 Clusters  & 47 & 46.48 & (44, 49)  \\ 
  Large Size $\geq$ 5 Clusters  & 41 & 40.76 & (38, 43)  \\ 
  Large Size $\geq$ 6 Clusters  & 35 & 34.81 & (33, 37)  \\ 
  \hline
\end{tabular}
\end{table}

The results from Table \ref{tab:fullimagesummary} match our earlier expectations based on our likelihood specifications. Recall in Section \ref{subsec:hyper} that our specification for the priors on the components of the cluster-specific covariance matrices led to small values of the variances and covariances. This structure is so that craters that are similar in location and diameter can be grouped together, but the likelihood of dissimilar craters being grouped together is small. The tight credible intervals thus reflect our beliefs about the cluster-specific covariances.

\begin{figure}[!b]
    \centering
    \includegraphics[width=385.7pt]{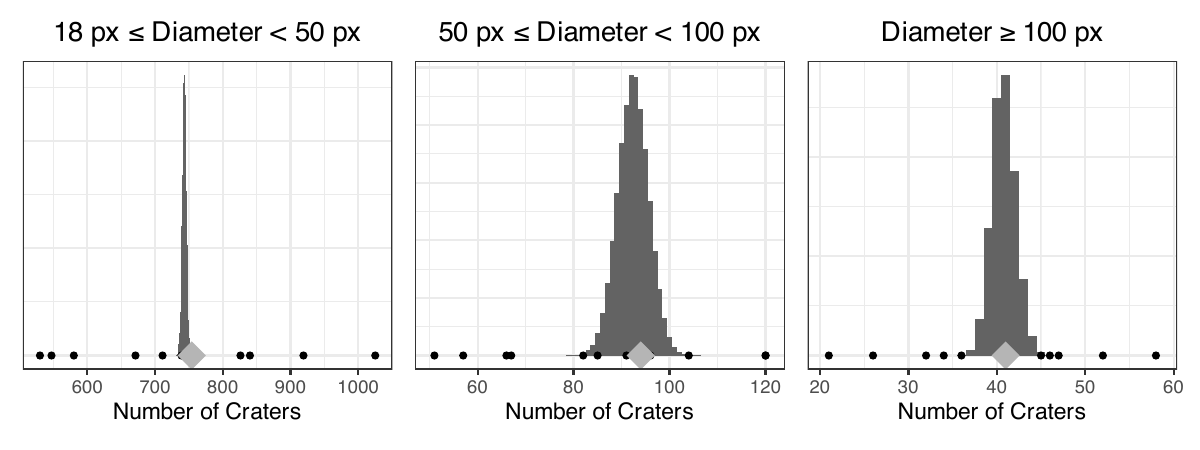}
    \caption{Number of small (left), mid-sized (middle), and large (right) craters identified by each expert (black circles), the DBSCAN algorithm (gray diamonds) used in R14, and DFCRP posterior consensus clusters (size $\geq$ 5; histogram). The DFCRP shows moderate agreement with DBSCAN.}
    \label{fig:uq}
\end{figure}

Figure \ref{fig:uq} compares the results for consensus (size $\geq$ 5) clusters identified by the DFCRP, with uncertainty, to the DBSCAN results of R14, and the individual experts separated by small, medium, and large craters. For all sizes, the variability across individual experts is appreciably larger than the variability of the consensus clusters found by the DFCRP. For instance, the range in the number of small craters identified by individual experts is between 530 and 1025 while the range from the DFCRP posterior draws is approximately between 731 and 755. The number of craters identified by the DBSCAN method is fairly consistent with the DFCRP for all crater sizes.

Draws from the posterior distribution of cluster assignments allow us to investigate expert-specific crater marking tendencies beyond the insights drawn from simple counts. For instance, we can use the posterior samples to provide insight into questions like:
\begin{enumerate}
\item How similar are experts?
\item Do some experts identify craters others fail to identify?
\item Do some experts fail to identify craters all other experts identify? 
\end{enumerate}

How similar are experts? To address this question, we computed the average Jaccard similarity coefficient between each pair of experts. The Jaccard similarity coefficient \citep{j01} is a measure of the intersection of two sets divided by their union. The measure is between 0 and 1, where 1 represents maximal similarity. Average pairwise Jaccard similarity coefficients for each expert are given in column 3 of Table \ref{tab:expert}, and a heatmap of the pairwise Jaccard similarity coefficients can be found in Section E of the supplementary material. In general, experts that found the most observational craters are the most similar to one another, while the most dissimilar experts are the ones that found the largest difference in the number of observational craters. The most dissimilar experts are experts D and A along with D and J. Expert D identified the most observational craters, while experts A and J identified the fewest observational craters.

This comparison of similarity between experts reveals a potential avenue for further research. The pairwise Jaccard similarity coefficients between experts reveal that experts H and I had the most observational craters clustered together. This is not surprising, as experts H and I are actually the same person using two different software systems to identify craters. Our current model has no method to incorporate the dependence between experts with shared characteristics (either the same scientist using different software or different scientists using the same software). It's reasonable to assume that craters identified by either the same crater counter or the same algorithm are more likely to be clustered together; one possible way to incorporate this dependency is to assign a higher probability to partitions that adhere to this structure, as seen in \cite{ddt17}.

Do some experts identify craters others fail to identify? To address this question, we identified each cluster of size 1 for all posterior cluster assignment draws and recorded the percentage of times an expert was in a cluster of size 1. The averages and 95\% credible intervals are presented in columns 4 and 5 of Table \ref{tab:expert}. Clusters of size 1 are interpreted as craters that one expert identified but all others failed to identify. These clusters may represent difficult to identify craters or incorrectly identified craters; however,  we leave that distinction for the crater counting community. Expert D made up roughly 50\% of size 1 clusters. This finding is not wholly surprising, as expert D identified over 100 more observational craters than any other expert. By virtue of finding more observational craters and the DFCRP explicitly taking into account the family structure of the data, we would expect the expert with the most identified observational craters to be in the most clusters of size 1. Expert B, however, made up roughly 15\% of all size 1 clusters, the second most, but did not identify the second most total observational craters. A figure with the location of the size 1 clusters identified by experts D and B is in Section E of the supplementary material. Interestingly, the size 1 clusters including expert B are predominantly isolated in the top part of the image, possibly suggesting that expert B's process of identifying moon craters differed there as compared to the rest of the image.

Do some experts fail to identify craters all other experts identify? To address this question, we identified all clusters of size 10 for each posterior sample and recorded the expert not included in the size 10 cluster. Size 10 clusters are craters that all but one expert identified. Averages and 95\% credible intervals are presented in columns 6 and 7 of Table \ref{tab:expert} for each expert. We see that, on balance, the fewer observational clusters an expert identified, the more likely they were to not identify a crater all other experts identified. One exception to this is expert J. Experts J and C both identified roughly the same number of observational craters, but expert J was almost 3 times more likely to not identify a crater all other experts identified. A similar finding holds when comparing expert F to expert E. They both identified a similar number of observational craters, but expert F was about 3 times more likely to not identify a crater all other experts identified than expert E.

\begin{table}[!tb]
\centering
\caption{Expert-specific clustering summaries ordered by the number of identified observational craters (count). Jaccard is the average pairwise Jaccard similarity coefficient, measuring similarity among experts. Size 1 is the average percentage of size-1 clusters in which each expert was included. Size 10 is the average percentage of size-10 clusters in which each expert was not included.}
\label{tab:expert}
\begin{tabular}{c|c|c|rc|rc}
  \hline
\textbf{Expert} & \textbf{Count} & \textbf{Jaccard} & \textbf{Size 1} & \boldmath{($95\% \textbf{ CI}$)}  & \textbf{Size 10} & \boldmath{($95\% \textbf{ CI}$)}  \\ 
  \hline
A & 636 & 0.62 & 0.422  & (0.256, 0.521) & 13.480  & (13.274, 13.636) \\ 
J & 652 & 0.59 & 0.777  &  (0.763, 0.785) & 37.693 & (36.937, 38.182) \\ 
C & 653 & 0.62 & 1.018  & (0.769, 1.295) & 13.547  & (13.393, 14.286) \\ 
F & 808 & 0.63 & 2.601  & (2.308, 3.101) & 18.225  & (17.857, 18.919) \\ 
E & 814 & 0.66 & 1.332 & (1.276, 1.554)  & 5.393   & (5.310, 5.455) \\ 
G & 831 & 0.65 & 1.565 &  (1.527, 1.809)    & 5.371   & (4.545, 5.455) \\ 
K & 930 & 0.64 & 7.834    &(7.494, 8.226)  & 2.697   & (2.655, 2.727) \\ 
B & 945 & 0.63 & 14.702 & (14.433, 14.948)    & 0.900   & (0.885, 0.909) \\ 
I & 991 & 0.64 & 10.252  &(10.000, 10.513) & 0.899  & (0.885, 0.909) \\ 
H & 1060 & 0.64 & 10.180   &(9.819, 10.513)      & 0.000   & (0.000, 0.000) \\ 
D & 1197 & 0.56 & 49.317  & (48.843, 49.742)    & 1.797   & (1.770, 1.818) \\ 
   \hline
\end{tabular}
\end{table}

\section{Conclusions and Remarks}
\label{sec:conclusions}
In this paper we presented a novel clustering approach to combining expert identified objects in an image. We incorporate a dysfunctional family constraint into the Chinese Restaurant Process (CRP) clustering approach, resulting in the dysfunctional family CRP (DFCRP). This constraint respects the number of objects identified by each individual expert by forcing objects identified by the same expert to be placed in different clusters. We have also explored some expert-specific considerations that could be addressed given draws from the posterior distribution of cluster assignments. This does not represent an exhaustive list of what is possible, but demonstrates one advantage of quantifying the uncertainty on cluster assignments. 

The development of the DFCRP was motivated by the work of R14, who explored the commonly held exchangeability of experts assumption for identifying craters on the moon. Their work showed that multiple experts identifying craters within the same image produced both crater identification and specification variability, implying that  conclusions based on a single image are potentially biased with underestimated precision. R14 combined the experts' opinions using clustering, through a DBSCAN modification. Our approach, the DFCRP, is also to cluster, with the additional ability to include an arbitrary number of features, quantify uncertainty on the cluster assignment, and  include the dysfunctional family constraint. 

The DFCRP was applied to the list of expert-identified craters for the lunar highlands image, and the results were similar to those obtained by R14. For this application, it is difficult to assess which method did ``better'', as the ground truth is unknown. Clearly, we believe the DFCRP is a more desirable approach based on the reasons listed above, but it is impossible to state which method gave a more accurate description of the craters on the moon. Our simulation study showed that, when family information is available, including the dysfunctional family constraint does lead to better performance over the CRP. 

This leads to an important distinction that should be made as to what the DFCRP does \textit{not} do. We do not claim to more accurately locate the true craters. This requires either  highly skilled subject-matter experts or well-trained algorithms, or knowledge of ground truth. Our framework does provide a mechanism to intelligently combine lists of identified objects from the same image  while naturally accounting for the structure of multiple identifications and quantifying uncertainty. Operations such as thresholding on the combined list, as was done in R14, is one possible way to define the true objects of interest. 

In this paper, we have applied this clustering approach to an object identification problem, but the methods presented in this paper apply to a wide variety of clustering problems in society today. For example, the DFCRP could be used to deduplicate hospital records for public health and research purposes. Alternatively, the DFCRP is a sensible choice for a problem such as election fraud between counties, ensuring that the same person is not registered to vote in multiple counties. The DFCRP is a natural fit to help businesses match which transactions are the same between two systems. This algorithm is also well-suited for genealogical work by identifying whether duplicate records are actually the same person. These examples illustrate the broader potential of the DFCRP across a wide range of structured clustering problems.\\

\noindent{\it \bf Significance:}
Scientists often rely on human experts to identify objects in images, such as craters on the Moon. However, different experts frequently disagree, leading to uncertainty in scientific measurements and conclusions. Existing methods for combining expert annotations do not fully account for this structured disagreement or provide principled measures of uncertainty. We propose a new statistical framework that combines multiple expert annotations while explicitly respecting the structure of how experts contribute to the data and quantifying uncertainty in the resulting object estimates. Applied to lunar crater identification, our method produces coherent groupings of observations that reflect both within- and between-expert variation. Our results show that accounting for expert-specific structure leads to more consistent aggregation of human annotations compared to standard clustering approaches. More broadly, this provides a principled way to combine multiple noisy human or algorithmic judgments in scientific image analysis, with implications for any setting where objects are identified subjectively by multiple observers.\\

\noindent{\it \bf Acknowledgments:} We would like to thank David Dahl, Stephen Vardeman, Ken Ryan, Mike Hamada, and Jim Gattiker for their helpful conversations. We would also like to thank C.C. Essix for her encouragement and support. This article approved for unlimited release (LA-UR-26-24138).\\
 
\bibliographystyle{agsm}
\bibliography{DFCRP} 

\newpage
\appendix

\noindent {\bf \LARGE Supplemental Material}

\bigskip
\noindent This supplementary material has five sections. The first section explains the Metropolis-Hastings sampler used to obtain draws from the conditional posterior distribution of $\alpha$. The subsequent section is in regards to the permutation parameter we implement to make our model order invariant (but not exchangeable). We then outline our procedure for generating the data used in the simulation study and provide the hyperparameters used. The fourth section provides further details on the reduced dataset used in Section 6.1 and the simulated dataset used in Section 7, and describes the priors utilized therein. Finally, we include additional figures relevant to the posterior analysis presented in Section 8. 

\section{\texorpdfstring{Prior on $\alpha$}{Prior on Alpha}}

The conditional posterior distribution of $\alpha$ is given by the following equation: 
\begin{equation*}
p(\alpha \mid \bc, \fmly)\propto p_{\dfcrp}(\bc \mid \alpha, \fmly)\,p(\alpha).
\end{equation*}
Since the data are modeled using the DFCRP, the distribution of cluster assignments $\bc$ conditional on $\alpha$ and the family assignment vector $\fmly$ takes the form of the joint distribution of table assignments, as in Equation 4.4. Because this distribution has an intractable normalizing constant, the posterior $p(\alpha \mid \bc, \fmly)$ is not available in closed form. We therefore employ a Metropolis-Hastings sampler rather than a Gibbs sampler to draw from the posterior distribution of $\alpha$, as shown in Algorithm \ref{alg:mhalpha}. 

We use a lognormal distribution as our proposal density to obtain $M$ draws from the conditional posterior $p(\alpha \mid \bc, \fmly)$. We define $q(\alpha^* \mid \alpha^{(m-1)})$ to be the density of a lognormal random variable with mean $\alpha^{(m-1)}$ and variance the inverse of a positive real number $\tau$; this corresponds to $\text{LN}(\mu = \ln(\alpha^{(m-1)})- 0.5\tau, \sigma = \tau^{-1/2})$. Thus, the mean of the proposal distribution is the current value of $\alpha^{(m-1)}$, but the variance can be controlled as necessary, allowing flexibility in the scale of the proposal. Throughout the paper, we used 100 for the  hyperparameter value of $\tau$.

We define $p(\alpha)$ as the prior distribution on $\alpha$. As specified in Section 5, we employ a $\text{Gamma}(a_{\alpha}, b_{\alpha})$ distribution for flexibility and ease in defining a prior applicable to a specific application. The values of $a_{\alpha}$ and $b_{\alpha}$ must be intentionally chosen to reflect the information known \textit{a priori}. 

\spacingset{1} 
\RestyleAlgo{boxruled} 
\LinesNumbered 
\begin{algorithm}[ht] 
    \caption{Metropolis-Hastings Sampler for $\alpha$\label{alg:mhalpha}} 
    \begin{algorithmic}[1] 
        \STATE Given initial values for $\alpha^{(m-1)}$, $\bc$, and $\bsigma$: 
        \STATE Choose: $\alpha^*\sim \text{LN}(\ln(\alpha^{(m-1)})- 0.5\tau, \tau^{-1/2})$.
        \STATE Calculate the acceptance ratio: \begin{equation*} 
        r=\frac{p_{\dfcrp}(\bc \mid \alpha^*, \fmly, \bsigma)\,p(\alpha^*)\,q(\alpha^{(m-1)} \mid \alpha^*)}{p_{\dfcrp}(\bc \mid \alpha^{(m-1)},\fmly, \bsigma)\,p(\alpha^{(m-1)})\,q(\alpha^* \mid \alpha^{(m-1)})}.
        \end{equation*} 
        \STATE Draw $u\sim\text{Unif}(0,1)$. 
        \STATE If $u \leq \text{min}(1, r)$ set $\alpha^{(m)}=\alpha^*$, else set $\alpha^{(m)}=\alpha^{(m-1)}$. 
    \end{algorithmic} 
\end{algorithm}
\spacingset{1.8} 

\section{Handling Permutation Dependence in the DFCRP} The following subsections provide details on the modifications made to the DFCRP to account for violated exchangability incurred upon incorporating expert structure into the model. We first provide an example of how the family structure violates the principle of exchangability. We then include details on how we sample from the posterior permutation space via a Metropolis sampler. Finally, we include support for the validity and effectiveness of our modification to the permutation proposal algorithm.

\subsection{A Demonstration of Order-Dependent Clustering}

To demonstrate the permutation dependence, we use an example which builds off the Chinese restaurant metaphor. Suppose we have three customers enter the restaurant, A, B, and C. Let the family assignments of the customers be $x_A=1$, $x_B=1$, and $x_C=2$. Additionally, assume that the distribution of tables follows a DFCRP with $\alpha=1$. Customer A is the first to enter the restaurant, and by so doing creates a new table with probability 1. Customer B enters next, which also creates a new table with probability 1, because $x_A=x_B$ (i.e., customers A and B are in the same family). Finally, customer C enters the restaurant, and (because there is no familial relation between it and the customers already in the restaurant) the probability distribution follows a standard CRP. Customer C creates a third table, which in this example has probability $\frac{1}{3}$. Thus, $p^*_{\dfcrp}(c_A=1, c_B=2, c_C=3 \mid \alpha=1, x_A=1, x_B=1, x_C=2) = \frac{1}{3}$.

Now assume that the permutation of the customers is different. Let the order in which they enter the restaurant be customer A, then customer C, and lastly customer B.  Again $c_A = 1$ with probability 1. Next, $c_C=2$ occurs with probability $\frac{1}{2}$.  Finally, $c_B=3$ has probability $\frac{1}{2}$. Thus the partition with each item in a separate cluster has probability: $p^*_{\dfcrp}(c_A=1, c_B=3, c_C=2 \mid \alpha=1, x_A=1,x_B=1,x_C=2) = \frac{1}{4}$.  This is the same partition as in the previous example, but it has a different probability! This demonstrates that the DFCRP is not exchangeable.

\subsection{\texorpdfstring{Marginalizing over $\bsigma$}{Marginalizing over Sigma}}

Summing out the permutation parameter becomes infeasible as $n$ increases, and this inefficiency grows with $n!$. Thus, to decrease computation time, we average over a small number of permutations of the $n$ items. Ideally, the selected permutations would be those with the greatest posterior probability. Thus, we use a Metropolis algorithm to identify permutations with high probability mass and average across those. We obtain $M$ samples from the conditional distribution $p(\bsigma \mid \bcmperm{n}, \alpha, \fmly)$ using the steps outlined in Algorithm \ref{alg:mhperm}. Note that we omit the prior on $\bsigma$ in the acceptance ratio because, in this work, $\bsigma$ is assigned a uniform prior.

\spacingset{1} 
\RestyleAlgo{boxruled} 
\LinesNumbered 
\begin{algorithm}[ht] 
    \caption{Metropolis Sampler for $\bsigma$\label{alg:mhperm}} 
    \begin{algorithmic}[1] 
        \STATE Given initial values for $\bsigma^{(i-1)}$, $\bc$, and $\alpha$:  
        \STATE Define a proposal distribution $q(\cdot \mid \cdot)$ for generating candidate permutations (i.e. random transposition, random insertion, etc.). See Section 4.2 for how we implement this in our sampler.
        \STATE Generate a candidate permutation, $\bsigma^*$ from $q(\bsigma^* \mid \bsigma^{(i-1)})$
        \STATE Calculate the acceptance ratio:         \begin{equation}\nonumber
    r=\frac{p{\dfcrp}(\bc \mid \alpha, \fmly, \bsigma^*)\,q(\bsigma^{(i-1)} \mid \bsigma^*)}{p{\dfcrp}(\bc \mid \alpha, \fmly, \bsigma^{(i-1)})\,q(\bsigma^* \mid \bsigma^{(i-1)})}
    \end{equation}
        \STATE Draw $u\sim\text{Unif}(0,1)$ 
        \STATE If $u \leq \text{min}(1, r)$ set $\bsigma^{(i)}=\bsigma^*$, else set $\bsigma^{(i)}=\bsigma^{(i-1)}$. 
    \end{algorithmic} 
\end{algorithm}
\spacingset{1.8} 

After generating $M$ permutations from this MCMC algorithm, we can approximate the process in Equation 4.6:
\begin{align*}
       p_{\dfcrp}(\bc  \mid  \alpha,\fmly) 
    &= \sum_{\bsigma \in \mathcal{S}} p_{\dfcrp}(\bc \mid \alpha, \fmly, \bsigma) \,p(\bsigma) \\
    &\approx \frac{1}{M}\sum_{m = 1}^M p_{\dfcrp}(\bc \mid \alpha, \fmly, \bsigma^{(m)}),
\end{align*}

where $\mathcal{S}$ is the set of all permutations of $n$ elements.

\subsection{The Permutation Proposal Modification}

As discussed in Section 4.2, the cost of running the sampler principally involves computation of the nonexchangable DFCRP prior. This takes roughly 5.88 milliseconds to compute the prior for a partition generated from the full image data. Note that for this paper, most calculations were conducted on a server featuring an Intel Xeon Platinum 8592+ CPU with 128 cores and 512 GB of memory. For the full dataset, this results in millions of prior evaluations per iteration, making the standard implementation computationally infeasible. Accordingly, we presented a modification to the Metropolis sampler for the permutation designed to make analysis of the full image data reasonable.

The idea of modifying the proposal mechanism for the permutation Metropolis sampler to reduce the number of times needed to compute the full DFCRP prior seems reasonable. We recognize, however, that empirical testing would provide proof of concept, and thus we compared the empirical results of the permutation modified sampler with theoretical values to evaluate the validity of the modification.

To generate empirical draws, we simulated a dataset with 6 observations and 2 families. After running the permutation-modified Gibbs sampler (with fixed $\alpha = 1$ and without the likelihood contribution) on the dataset for 50,000 iterations and thinning to keep only 1 out of every 10 draws, we computed the proportion of each partition from the posterior draws. For the theoretical values, we computed the full DFCRP prior for each permutation and each partition allowed for our simulated dataset (i.e. all the partitions in the empirical results). In other words, we computed a matrix $\boldsymbol{X}$ where 
$$ x_{ij} = P_{DFCRP}(\bc_j \mid \bsigma_i, \alpha = 1,  \fmly).
$$

\noindent We then marginalized out the permutation parameter $\bsigma$ by taking the column means for the matrix, and compared the resulting probabilities to the empirical values. The results are in Table \ref{tab:permmod}. We found that none of the empirical proportions had a difference of greater than 0.01 when compared to the true values. Additionally, a chi-square goodness-of-fit test on these proportions yielded a $p$-value of 0.2398. We accordingly fail to reject the null hypothesis and conclude that the permutation-modified Gibbs sampler is a valid sampling algorithm.

\begin{table}[ht]
\caption{Empirical and theoretical values for the 21 valid partitions of a simulated dataset with 6 observations and 2 families. 'Empirical' indicates the proportion of draws out of 5,000 that resulted in the indicated partition. 'Theoretical' denotes the theoretical proportion of the partition under the DFCRP prior, and 'Difference' is the difference between the two proportions.}
\centering
\begin{tabular}{rrrrrrrr}
  \hline
Partition & Empirical & Theoretical & Difference & Partition & Empirical & Theoretical & Difference \\ 
  \hline
  1 & 0.072 & 0.062 & -0.010 & 
  2 & 0.063 & 0.062 & -0.001 \\ 
  3 & 0.064 & 0.062 & -0.001 &
  4 & 0.029 & 0.030 & 0.001 \\ 
  5 & 0.060 & 0.062 & 0.002 &
  6 & 0.065 & 0.062 & -0.003 \\ 
  7 & 0.057 & 0.062 & 0.006 &
  8 & 0.030 & 0.030 & -0.000 \\ 
  9 & 0.058 & 0.062 & 0.004 & 
  10 & 0.062 & 0.062 & -0.000 \\ 
  11 & 0.061 & 0.062 & 0.002 &
  12 & 0.027 & 0.030 & 0.003 \\ 
  13 & 0.061 & 0.062 & 0.001 &
  14 & 0.063 & 0.062 & -0.001 \\ 
  15 & 0.066 & 0.062 & -0.004 &
  16 & 0.033 & 0.030 & -0.003 \\ 
  17 & 0.030 & 0.030 & -0.000 &
  18 & 0.026 & 0.030 & 0.004 \\ 
  19 & 0.028 & 0.030 & 0.002 &
  20 & 0.027 & 0.030 & 0.003 \\ 
  21 & 0.017 & 0.015 & -0.002 &
     &       &       &       \\
\hline
\end{tabular}
\label{tab:permmod}
\end{table}

To measure the effect of this modification on the runtime of the sampler, we computed the effective sample size (ESS) of the average of the first half of the indices in the permutation. This measure of the state of the permutation is continuous, and thus a measure of the ESS is much more representative of the mixing of the permutation. We ran 30 chains of the Gibbs sampler both with and without our permutation modification, and measured the time it took to obtain an ESS of 5 for the averaged first half of the permutation. Across these 30 chains, the average time required without the permutation modification was 173.8 minutes. With the permutation modification, this dropped to 18.7 minutes, or a 9.3x speedup. Consequently, we conclude that modifying the permutation proposal algorithm in this way is both valid and computationally more efficient.

\section{Simulated Data Generation}

In Section 7, we simulated 500 datasets according to Algorithm \ref{alg:simdat}, where $K=30$, $L_X = L_Y = 0$, $U_X = 700$, $U_Y = 500$, $a_{lD} = 64$, $b_{lD} = 16$, $J=6$, $\boldsymbol{t} = (0.98,0.96,0.94,0.92,0.9,0.88)$, $\boldsymbol{f} = (0.12, 0.1, 0.08, 0.06, 0.04, 0.02)$, and $\bSigma = \text{diag}(5, 5, 0.01)$. We chose these parameters such that the features of the simulated dataset were similar to the reduced dataset used in Section 6.1.

\spacingset{1} 
\hbadness=10000
\RestyleAlgo{boxruled}
\LinesNumbered
\begin{algorithm}[ht]
  \caption{Procedure for Data Generation\label{alg:simdat}}
  \begin{algorithmic}[1]
  \STATE Choose $K_t$ to be the number of true clusters for the simulated dataset, and J to be the number of experts analyzing the data.
  \FOR{$k$ in $1,\ldots,K_t$}
    \STATE Draw $X_k$ from Uniform(L$_X$,U$_X$)
    \STATE Draw $Y_k$ from Uniform(L$_Y$,U$_Y$)
    \STATE Draw $lD_k$ from Gamma($a_{lD},b_{lD}$)
    \STATE Set $\bmu_k = (X_k,Y_k,lD_k)'$
    \STATE Set $\bSigma$ equal to a user-specified covariance matrix
    \ENDFOR
\FOR{$j$ in $1,2,\ldots,J$}
    \STATE Sample which clusters $k_t$ expert $j$ will identify using a predetermined true cluster detection vector $\boldsymbol{t}$
    \FOR{$k$ in $k_t$}
        \STATE Draw $\by_{j,k}$ from N($\bmu_k,\bSigma$)
        \STATE Set $ \fmly_k[k] = x_{i,k}$, where $x_{i,k}$ is the family assignment for $\by_{i,k}$. $\fmly_k$ constitutes the \newline family assignment for the $n_k$ data points assigned to cluster $k$.
        \STATE Set $z_{i,k} = k$, the cluster assignment for $\by_{i,k}$
    \ENDFOR
    \STATE Sample how many false clusters $K_f$ expert $j$ will identify using a predetermined false cluster \newline detection vector $\boldsymbol{f}$
    \FOR{$k$ in $K_t+1,\ldots,K_t+K_f$}
        \STATE Draw $X_k$ from Uniform(L$_X$,U$_X$)
        \STATE Draw $Y_k$ from Uniform(L$_Y$,U$_Y$)
        \STATE Draw $lD_k$ from Gamma($a_{lD},b_{lD}$)
        \STATE Set $\by_{j,k} = (X_k,Y_k,lD_k)'$
        \STATE Set $x_{i,k} = \fmly_k[k]$
        \STATE Set $z_{i,k} = k$, the cluster assignment for $\by_{i,k}$
    \ENDFOR
\ENDFOR
\end{algorithmic}
\end{algorithm} 
\spacingset{1.8} 

\section{Details on the Reduced Dataset}

In Section 6 we utilize a reduced dataset to test clustering principles without expending the computation required to examine the full dataset. We first use a reduced dataset to test different values of the neighborhood radius $\rho$. For this application, we used only craters within the $x$-coordinates $(1700, \; 2400)$ and within the $y$-coordinates $(-300, \; 0)$. This left us with a rectangular region at the top center of the image. The only modifications to the model that were necessary for this smaller dataset was to specify new values of $\bmu_0$ and $\bSigma_0$, the prior on the cluster-specific means $\bmu$ and the prior on $\alpha$. By obtaining draws from the prior and comparing them with the craters present in the reduced dataset, we determined that $\bmu_0 = (2050,\;  -150, \; 3.2)$ and $\bSigma_0 = \text{diag}(200^2, \; 110^2, \; 0.5^2)$ specified a prior that was appropriate for our application. Due to the reduced size of the dataset, it was also necessary to reevaluate the specification for the prior on $\alpha$. We set $a_{\alpha} = 1$ and $b_{\alpha} = 0.01$, producing a prior that favors lower cluster counts, which is appropriate given the size of the reduced dataset.

In comparing the DFCRP with the CRP in our simulation study, we chose parameters for the dataset that would resemble the reduced dataset used in Section 7 to test the radius parameter $\rho$. Accordingly, we again took draws from the prior $p(\bmu)$ with different specifications for $\bmu_0$ and $\bSigma_0$, and found that $\bmu_0 = (350, 250, 3.9)$ and $\bSigma_0 = \text{diag}(300^2, 225^2, 0.45^2)$ generated draws that were comparable to the craters in the simulated dataset.

\section{Figures}

Below is the Jaccard Similarity Coefficient heatmap referenced in Section 8. Also below is the plot of all craters of size 1 from experts D and B, the experts with the most observational craters. That image is also referenced in Section 8.

\begin{figure}[!b]
    \centering
    \includegraphics[width=385.7pt]{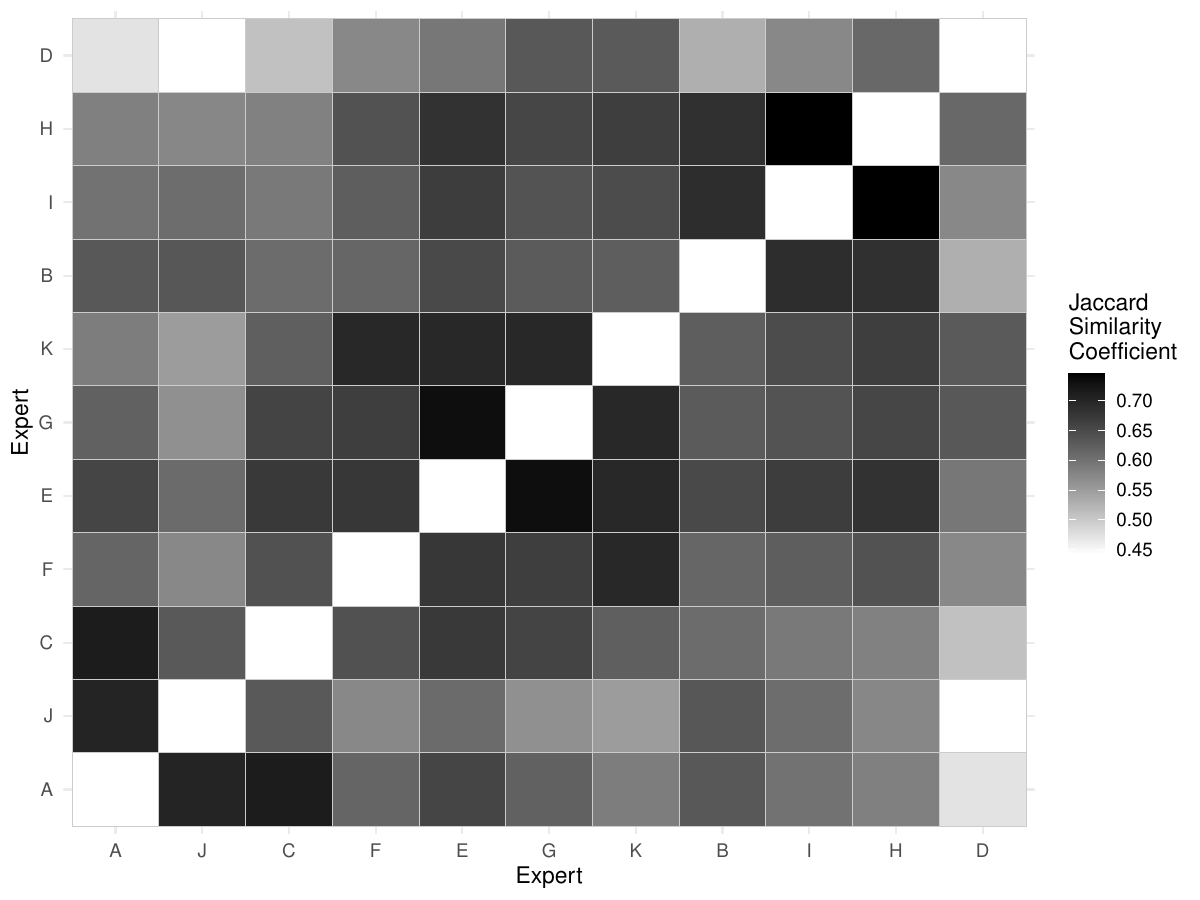}
    \caption{Pairwise Jaccard similarity coefficients. Experts are ordered by number of identified observational craters, from left to right and bottom to top. Darker shades indicate more similarity between the craters identified by two experts.}
    \label{fig:jaccard}
\end{figure}

\begin{figure}[!tb]
    \centering
    \fbox{\includegraphics[width=385.7pt]{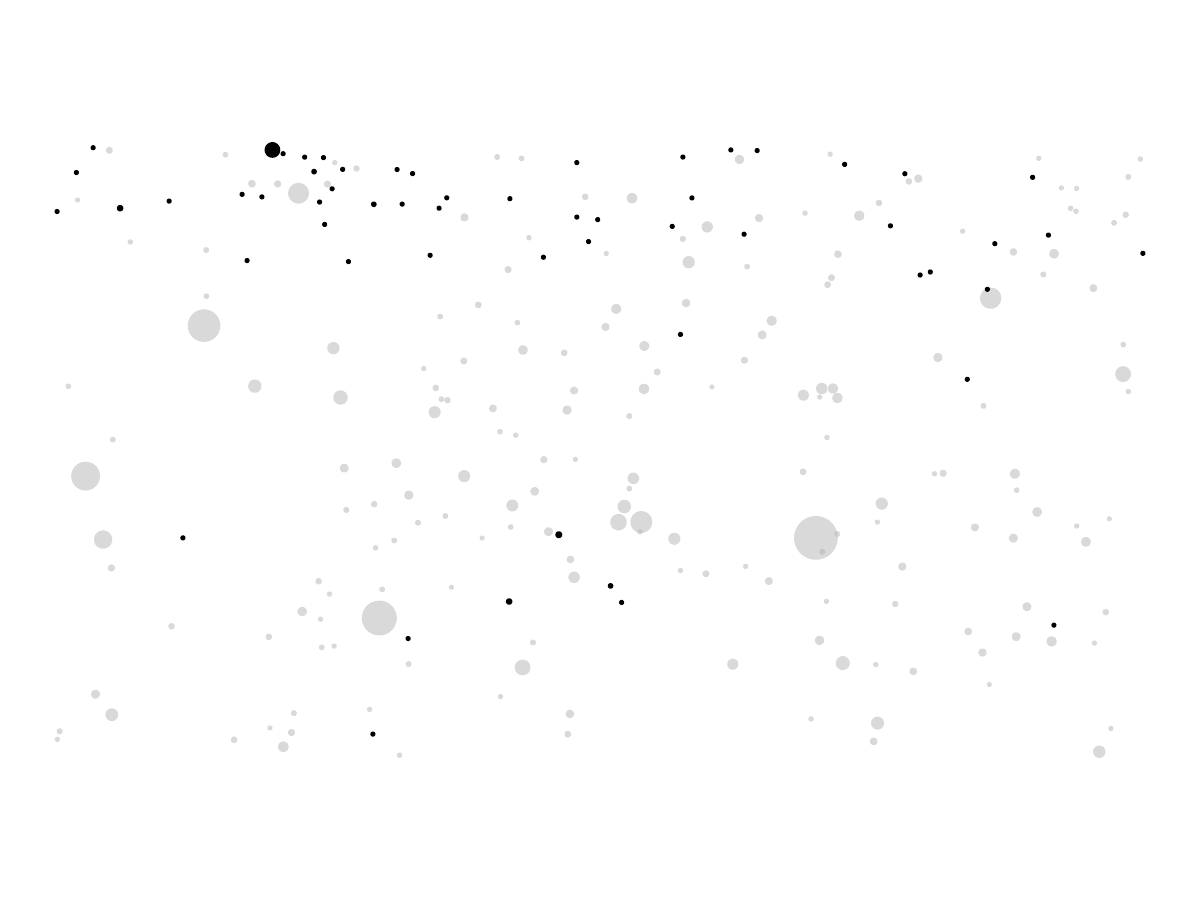}}
    \caption{All observational craters for the lunar highlands image that were size 1 clusters containing expert D (grey) and expert B (black). Expert D was included in roughly 50\% of all size 1 clusters, and expert B was included in about 15\% of those clusters. Note that the size 1 clusters of expert B were concentrated in the upper portion of the image.}
    \label{fig:size1}
\end{figure}

\end{document}